\documentclass[%
 reprint,
 superscriptaddress,
 twocolumn,
 amsmath,amssymb,
 aps,
prx,
floatfix,
]{revtex4-2}
\usepackage{graphicx}% Include figure files
\usepackage{dcolumn}% Align table columns on decimal point
\usepackage{bm}% bold math
\usepackage{siunitx}
\usepackage{orcidlink}
\usepackage{array}
\usepackage{multirow}

\RequirePackage{xcolor}
\definecolor{trailer}{gray}{0.9}

\def\formtmp#1#2{{\vskip6pt\noindent\fboxsep=1pt\colorbox{#1}{\vbox{\vskip1pt\hbox to \columnwidth{\hskip3pt\vbox{\raggedright\noindent\textbf{#2\vphantom{Qy}}}\hfill}\vspace*{0pt}}}\par\vskip0pt%
\noindent\kern0pt}}

\newenvironment{trailer}[1]{\ignorespaces\def\stmtopen##1{##1}%
\formtmp{trailer}{#1}}{\par\noindent\textcolor{trailer}{\rule{\columnwidth}{1pt}}\vskip0pt\par\addvspace{\baselineskip}}%

\def\Rb87{^{87}\mathrm{Rb}}                             % Rb 87
\def\K40{^{40}\mathrm{K}}                    		    % K 40 
\def\stimes{\!\times\!}                                 % short times

\def\OD{\mathrm{OD}} 
\usepackage{import}

\usepackage[columnwise]{lineno}
\begin{document}
%%%%%%%%%%%%%%%%%%%%%%%%%%%%%%%%%%%%%%%%%%%%%%%%%%%%%%%%%%%%%
\title{Can machine learning for quantum-gas experiments be explainable?}% Force line breaks with \\

\author{I.~B.~Spielman\orcidlink{0000-0003-1421-8652}}
\email{ian.spielman@nist.gov}
\affiliation{National Institute of Standards and Technology, Gaithersburg, MD 20899, USA}
\affiliation{Department of Physics, University of Maryland, College Park, MD 20742, USA}
\affiliation{Joint Quantum Institute, University of Maryland, College Park, MD 20742, USA}

\author{Justyna~P.~Zwolak\orcidlink{0000-0002-2286-3208}}
\email{jpzwolak@nist.gov}
\affiliation{National Institute of Standards and Technology, Gaithersburg, MD 20899, USA}
\affiliation{Department of Physics, University of Maryland, College Park, MD 20742, USA}
\affiliation{Joint Center for Quantum Information and Computer Science, University of Maryland, College Park, MD 20742, USA}

\date{\today}
%%%%%%%%%%%%%%%%%%%%%%%%%%%%%%%%%%%%%%%%%%%%%%%%%%%%%%%%%%%%%
\begin{abstract}
Virtually all aspects of many-body atomic physics are challenging: experiments are technically demanding, datasets have become enormous, and the memory and CPU requirements for classical simulation of generic quantum systems often scale exponentially with system size.
Machine learning (ML) methods are already assisting in each of these areas and are poised to become transformative.
Here, we focus on two specific applications of ML to cold-atom-based quantum simulators.
These devices generally generate data in the form of images; we first showcase denoising of raw images and then identify solitonic waves in Bose-Einstein condensates.
In both of these examples, we comment on the interplay between performance, model complexity, and interpretability.
\end{abstract}

\maketitle
%%%%%%%%%%%%%%%%%%%%%%%%%%%%%%%%%%%%%%%%%%%%%%%%%%%%%%%%%%%%%
%\pacs{}% insert suggested PACS numbers in braces on next line
%%%%%%%%%%%%%%%%%%%%%%%%%%%%%%%%%%%%%%%%%%%%%%%%%%%%%%%%%%%%%
%\linenumbers
%%%%%%%%%%%%%%%%%%%%%%%%%%%%%%%%%%%%%%%%%%%%%%%%%%%%%%%%%%%%%
\section{Introduction}
%%%%%%%%%%%%%%%%%%%%%%%%%%%%%%%%%%%%%%%%%%%%%%%%%%%%%%%%%%%%%
Today's atomic quantum simulators and computers are experimentally elaborate~\cite{Bruzewicz2019, Henriet2020}, produce increasingly large datasets, and are computationally bottlenecked: classical simulation of generic quantum systems scales exponentially with system size~\cite{Harrow2017}.
Machine learning (ML) methods already help navigate this combination of experimental complexity, data volume, and computational intractability, and are poised to have a transformational impact~\cite{Carleo2019}.
The preceding chapter by H.~Schl\"omer and A.~Bohrdt provides an up-to-date overview of the application of ML techniques to cold-atom physics, with both experimental and theoretical use cases.
In this chapter, we take a complementary tack by explicitly showcasing the use of ML for two specific experimental tasks: the construction of ``reference images'' used in the first step of experimental data processing~\cite{Ketterle1999}, and the identification of solitary waves~\cite{Denschlag2000, Carretero2008} in images of Bose--Einstein condensates (BECs).
Together, these highlight the importance of high-quality training data~\cite{Zha2025} and the trade-offs in model architecture between performance and scientific transparency~\cite{Rudin2019}. 

%%%%%%%%%%%%%%%%%%%%%%%%%%%%%%%%%%%%%%%%%%%%
\begin{figure*}[t]
\raggedright
\begin{minipage}[t]{0.33\linewidth}
\vspace{-5pt}
    \caption{
    Bright-field imaging for cold atoms.
    (a) Schematic of the imaging system.  
    A dust mote scatters light near the object plane, which is refocused near the image plane.
    (b) Raw image data showing a complex combination of interference patterns. 
    The left panel shows the probe with atoms present, the center panel shows the probe alone, and the right panel shows the background image.
    The red curves denote the reconstruction window.
    (c) Optical depth.  The left panel uses no reference recovery; the middle panel uses standard linear-algebra techniques; and the right panel uses a ResNet ML model.}
    \label{fig:imaging}
\end{minipage}
\begin{minipage}[t]{0.60\linewidth}
\vspace{0pt}
    \includegraphics{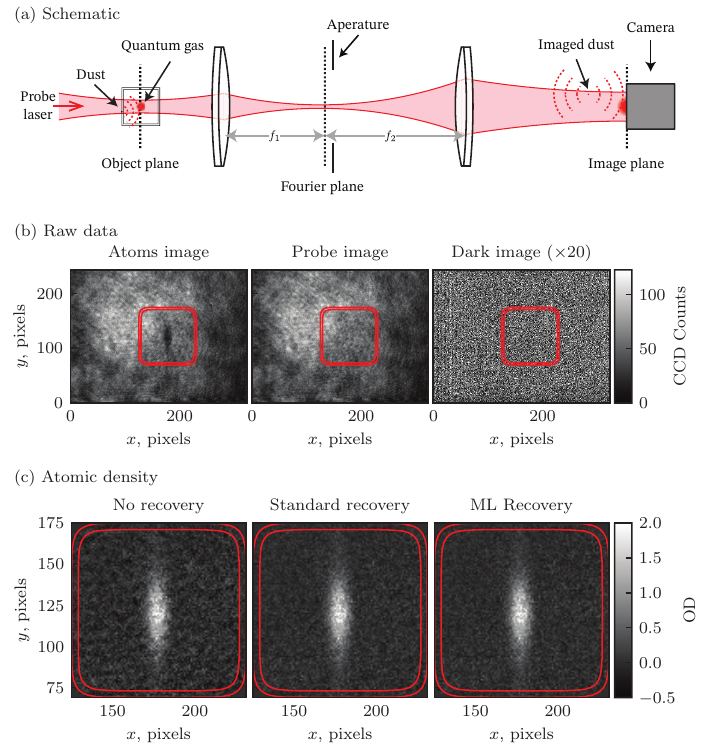}
\end{minipage}
\end{figure*}
%%%%%%%%%%%%%%%%%%%%%%%%%%%%%%%%%%%%%%%%%%%%

Almost all neutral atom quantum devices operate in a cyclic manner~\cite{Ketterle1999, Henriet2020}.
Atoms are first laser cooled from tens to thousands of Kelvin and captured in a magneto-optical trap (MOT).
A carefully scripted control sequence then performs further cooling and state preparation to yield the atomic ensemble under study.
After a series of classical control steps and one or more measurements, the ensemble is discarded, and the process repeats.
Thus, unlike solid-state quantum devices (and, to a lesser extent, ion-based systems), neutral-atom quantum systems are short-lived.
Every stage of this process can be improved using ML, including hardware design, control sequence discovery~\cite{Vendeiro2022}, measurement~\cite{Impertro2023}, and data processing~\cite{Bohrdt2019, Kaming2021}. 

Quantum devices offer new capabilities for sensing weak signals, simulating quantum phenomena, and performing a range of computational tasks~\cite{Degen2017, Altman2021, Alexeev2021}; however, their performance is often limited by measurement.
For example, in neutral-atom and ion arrays, the overall speed is limited by measurement times~\cite{Henriet2020}; atomic clocks and magnetometers require sub-Poissonian atom-number detection for quantum-enhanced sensing~\cite{Pezze2018}; and in cold-atom quantum simulators, the identification of quantum correlations is often limited by technical noise~\cite{Parsons2016}.
Most neutral-atom sensors, simulators, and computers acquire data as images; in particular, quantum gas experiments often rely on bright-field techniques with coherent laser light, such as absorption or phase-contrast imaging~\cite{Ketterle1999}, schematically illustrated in Fig.~\ref{fig:imaging}a.
As seen in Fig.~\ref{fig:imaging}b, coherent-light imaging such as this presents unique challenges because each unwanted source of optical scattering, for example, dust on optical elements or etaloning, adds its own interference pattern to the detected image.
The first section of this chapter contrasts traditional linear-algebra and ML techniques for removing these artifacts (Fig.~\ref{fig:imaging}c).
Processing such as this is essential, as each stage of the overall analysis pipeline impacts the next: for example, improvements in initial data processing yield higher-quality datasets, which benefit all downstream analysis and enable further refinements of the initial stage.

Even with high-quality denoised data, extracting a desired signal can be challenging.
Our second section demonstrates this stage of data analysis through the feature identification task of locating excitations, called solitons, in BECs~\cite{Denschlag2000}.
Solitons, robust shape-preserving solitary waves, are present in many physical systems and arise from an interplay between nonlinearity and dispersion.
In experiments with ultracold gases, solitonic excitations provide a vivid window into nonlinear, nonequilibrium physics, and (excluding systems where the spin degree of freedom is relevant) are either ``bright'' or ``dark''.
Bright solitons are stable for attractively interacting systems, where the tendency for the BEC to collapse to a single point is counterbalanced with the kinetic energy resulting from bending the macroscopic wave function.
The reverse holds for dark solitons, which are stable in repulsively interacting systems.
Owing to the comparative ease of preparing large repulsively interacting BECs, experiments with dark solitons are more common and are the focus of this chapter. 
Recent experimental developments have enabled the preparation of single solitons, or ensembles of solitons, with key properties---such as position and velocity---tunable by experimental design~\cite{Theocharis2010, Fritsch2020}.

In elongated BECs, dark solitons often appear as narrow density depletions in absorption or phase-contrast images, with physical information such as velocity and local background density reflected in the soliton depth and width.
Large collections of images acquired across experimental parameters can therefore reveal soliton statistics, formation mechanisms, and interaction dynamics.
In practice, however, much of this physics is locked inside high-volume image datasets.
A first obstacle is surprisingly basic: before we can study soliton physics quantitatively, we must reliably determine \textit{which images contain solitonic excitations}, \textit{the number of excitations}, and \textit{their positions}.
This ``localization-and-classification'' step quickly becomes the dominant bottleneck when experiments produce thousands to tens of thousands of images~\cite{Guo2021, Guo2022}.

The core difficulty in localizing and classifying solitonic excitations is the lack of a fixed, rigid visual template for ``a soliton in a BEC image.''
The observed depth and width of a dark soliton depend on its velocity and the imaging resolution, while shot-to-shot atom-number fluctuations, thermal effects, and imaging artifacts can create soliton-like dips or obscure genuine excitations.
The situation becomes even more challenging when multiple excitations coexist: solitons can be close together, merge, decay, or move between frames, complicating the implementation of simple rule-based detectors.
These realities make naive automated heuristics brittle.

One effective response is to use convolutional neural networks (CNNs) trained directly on images. 
CNNs can learn robust visual features for classification (e.g., ``no soliton'' versus ``soliton present'') and can be extended to localization or object detection when the task demands soliton positions. 
CNN-based approaches have been shown to dramatically speed up dark-soliton detection and enable high-throughput analysis that would be impractical by eye~\cite{Guo2021}.

CNNs introduce two practical challenges for scientific use.
First, as supervised models, they require substantial labeled data for training and validation; in the absence of reliable simulations, labels often come from manual annotation of experimental image collections, which is slow, labor-intensive, and can be inconsistent across annotators or across experimental conditions~\cite{Fritsch2022}.
Second, even when accuracy is high, CNN decision rules are typically opaque: the model may memorize subtle correlates (e.g., imaging artifacts or background structure) that are not part of the underlying soliton physics, complicating verification and failure-mode analysis.
Common explanation tools for CNNs, such as saliency maps~\cite{Simonyan2013} or Gradient-weighted Class Activation Mapping (Grad-CAM)~\cite{Selvaraju2020}, provide heuristic visualizations rather than a transparent model of \textit{how} each physical feature influences the decision. 
This is a critical issue when the goal is not just a label, but a trustworthy scientific inference workflow that includes an interpretable link between the label and the underlying physically meaningful cues~\cite{Guo2022}.

In this spirit, we consider alternatives---with CNN-level performance---that yield human-readable explanations of the evidence used for each decision.
Specifically, we compare a high-accuracy image-native CNN classifier with an explainable boosting machine (EBM) model that operates on physically motivated, low-dimensional image features and produces transparent, additive explanations~\cite{Lou2013}.
EBMs are a family of so-called \textit{glass-box} models: their predictions decompose exactly into a sum of learned feature-shape functions (and, optionally, a small number of pairwise interaction terms), enabling global visualizations and per-image attributions without post-hoc explanation methods~\cite{Nori2019}.
In our context, the fully intelligible EBMs can not only match but actually surpass CNN classifiers' performance, addressing the ``what did the model decide?'' as well as the ``what evidence did it use, and does that evidence align with the physics?'' questions.

With this backdrop, we can now articulate the most essential message of this chapter: ML works best starting with large quantities of high-quality data, a well-defined figure of merit, and the simplest viable model.
Data is the bedrock of ML: a good dataset (labeled or unlabeled) contains a dense and complete set of ``examples'' so that the right trained model can learn to interpolate between them~\cite{Zha2025}.
This underscores the importance of understanding the dataset's content: what are typical signals?  
What is the scale of noise or uncertainty?  
What features of the data are of interest?  
How often do these features appear?
The answers to this sort of question will guide the assembly and curation of the dataset.
ML is not a free lunch: good performance without spending considerable effort assembling the dataset is rare.

The ``complexity'' of an ML model depends on a range of interrelated considerations, including the size of the data vector traversing the model, the sophistication of the model architecture, the hyper-parameters required to describe the model (such as the number of layers in a neural network), as well as the overall number of parameters in the model.
However, in practice, the most common failure mode in applying ML to experimental physics is not an exotic choice of architecture, but an ill-posed learning objective: before collecting data or training a model, one must specify \textit{what} is to be learned and \textit{how success will be measured}.
This means translating a scientific goal into a concrete prediction task (classification, regression, denoising, segmentation), defining the target quantity and its domain (e.g., a ``reference image'' vs.\ an artifact-free density or excitation count vs.\ excitation locations vs.\ excitation type), and choosing labels or surrogate targets that are consistent with the physics and the measurement process.
Equally important is deciding what variability the model should be \textit{invariant} to (shot-to-shot atom number, global translation, fringe patterns) and what variability it should remain \textit{sensitive} to (true excitations, physically meaningful contrast changes), since these choices determine whether the model learns the intended signal or a convenient correlate.
Finally, a well-defined figure of merit turns ``it seems to work'' into a quantitative statement and exposes failure modes early.
As we will see, these choices influence training efficiency, performance, and interpretability.

%%%%%%%%%%%%%%%%%%%%%%%%%%%%%%%%%%%%%%%%%%%%%%%%%%%%%%%%%%%%%
\section{Experimental imaging of cold-atom systems}\label{sec:imaging}
%%%%%%%%%%%%%%%%%%%%%%%%%%%%%%%%%%%%%%%%%%%%%%%%%%%%%%%%%%%%%
Our examples employ bright-field techniques such as absorption or phase-contrast imaging. 
As depicted in Fig.~\ref{fig:imaging}a, an illuminating probe laser beam is first attenuated or phase-shifted by the atomic sample~\cite{Ketterle1999, Altuntas2021}; then, the resulting optical field traverses the imaging system and the final 2D intensity $I(x,y)$ is imaged as a function of position $(x,y)$.
Relevant atomic properties, such as the spatial density, are encoded in the fractional transmission $\simeq I(x,y) / I_{\rm p}(x,y)$, and therefore require knowledge of the probe intensity $I_{\rm p}(x,y)$ in the absence of atoms.
In practice, reference images are acquired shortly before or after the data image; during this interval, coherent-imaging artifacts often change slightly, leading to imperfect division.
For example, in Fig.~\ref{fig:imaging}a, the probe laser is shown to travel into a vacuum system (square) where it encounters a dust mote, leading to an outgoing scattered optical field.
This pattern of light is faithfully imaged to the ``imaged dust'' just prior to the camera, where the interference pattern between the scattered light and the probe, as attenuated by the atomic ensemble, is measured.
In realistic imaging systems, several such sources can be simultaneously present, leading to a highly structured probe laser, as in Fig.~\ref{fig:imaging}b (left).
Mitigating these artifacts is the first focus of this chapter, in Sec.~\ref{sec:ref_recon}.

In what follows, we exclusively employ resonant absorption imaging to obtain the 2D column density of atoms
\begin{align}
    n(x,y) &= \int dz\ \rho(x,y,z),
\end{align}
the 3D density $\rho(x,y,z)$ integrated along the imaging line of sight.
In the simplest case of low intensity, each atom contributes equally to the attenuation of the probe beam.
This leads to Beer's law exponential behavior characterized by the optical depth
\begin{align}\label{eq:OD}
    \OD(x,y) &= -\log\left(\frac{I(x,y) - I_{\rm bg}(x,y)}{I_{\rm p}(x,y) - I_{\rm bg}(x,y)}\right),
\end{align}
giving
\begin{align}
    n(x,y) &= \frac{\OD(x,y)}{\sigma_0},
\end{align}
where $\sigma_0 = 3 \lambda^2 / (2\pi)$ is the resonant absorption cross section for a probe laser of wavelength $\lambda$, and $I_{\rm bg}(x,y)$ is an additional image taken with the probe laser off to account for any common-mode background signal.
For brevity, this discussion omits several considerations that are irrelevant for reference reconstruction, but are important for accurate determination of $n(x,y)$; see Refs.~\cite{Ketterle1999} and ~\cite{Altuntas2021} for details.
It is common to release the atomic ensemble from its confining potential for a brief time-of-flight (TOF) prior to imaging.
In many cases, this period of nominally ballistic evolution enables direct access to the momentum distribution; however, for the data presented in Sec.~\ref {sec:solitons}, TOF expansion is dominated by interaction effects.
In most experiments, dark solitons are too small to be resolved in-situ; they are made visible by TOF expansion.

Figure~\ref{fig:imaging}b shows an example with atoms (left), probe-only (center), and dark (right) images; these data were taken in situ (i.e., with no TOF).
These data were selected to illustrate the consequences of a highly contaminated probe laser: numerous overlapping interference structures are visible.
When the scattering sources leading to this pattern spatially drift by even a fraction of the optical wavelength between these exposures (for example, due to acoustic vibrations of optical elements), coherent-imaging artifacts change noticeably, leading to imperfect division.
The red squircle (a function that interpolates between a square and a circle) encloses the expected location of the absorption signal, which is barely visible to the eye.
Simply computing the OD using Eq.~\eqref{eq:OD} yields the image in Fig.~\ref{fig:imaging}c (left) with artifacts comparable in magnitude to the underlying signal.
The artifacts are greatly reduced using both conventional linear algebra (middle) and ML methods (right), finally revealing the pattern of four separate groups of atoms.
These data were taken from Ref.~\cite{Zhao2025}, which followed the dynamics of these groups as they moved in a larger background BEC (not visible).

%%%%%%%%%%%%%%%%%%%%%%%%%%%%%%%%%%%%%%%%%%%%%%%%%%%%%%%%%%%%%
\section{Reference reconstruction}\label{sec:ref_recon}
%%%%%%%%%%%%%%%%%%%%%%%%%%%%%%%%%%%%%%%%%%%%%%%%%%%%%%%%%%%%%
With this basic understanding of bright-field imaging's how and why as it applies to cold-atom experiments, we return to the reference recovery task in more detail.
At a high level, the task is to take data as in Fig.~\ref{fig:imaging}b and predict the intensity $I_{\rm p}(x,y)$ that actually illuminated the atomic ensemble; in practice, this amounts to using the intensity exterior to the red squircle (where no atoms are present) to predict the interior intensity.
This can be accomplished by collecting a large ``training set'' of reference images (no atoms present) that reflect the full variability present in the probe beam.
For relatively clean beams, this might require 10 to 100 images; the depicted probe beam is a worst-case scenario, requiring a training set of $\approx 10^4$ images.
Using these data, both linear-algebra (LA) and ML techniques can be employed to construct an optimal reference image for each image containing atoms~\cite {Li2007a, Segal2010, Niu2018}.

LA methods are established tools that greatly reduce coherent-imaging artifacts, but because the observed intensity is a quadratic function of the underlying complex field, intensity-domain summation can require many linear combinations to eliminate residual artifacts.
When framed in the ML context, these are very simple, high-parameter-count models that lead to a highly compressed intermediate vector representation, i.e., a latent space.
In this context, such models are highly interpretable, perform well, and are trivial to ``train.''
Here, we contrast LA with more traditional ML techniques and explore the trade-offs between interpretability, performance, and development effort.

LA methods operate in a highly intuitive way.
They construct a collection of independent templates for the interior and exterior regions; then, for an image with atoms, they use the exterior-region templates to determine the appropriate linear combination of interior-region templates.
In this way, LA templates can have the physical interpretation patterns that tend to change together, ideally from individual scattering sources.
This physical interpretation breaks down for non-linear contributions resulting from interference between multiple sources.
A key secondary benefit of this approach is noise filtering: white noise from all sources (photon/photo-electron shot noise, read noise, ...) will generally have little in common with the templates and will be removed from the reconstructed reference.

By contrast, published ML-based tools for reference recovery use U-Net architectures where down-sampling CNN layers are followed by up-sampling with ConvTranspose or PixelShuffle layers, along with bypass connections~\cite{Ness2020, Ying2023, Lee2025}.

These models only slightly outperform LA approaches, are opaque black-box tools, and offer no in-principle reduction of white detection noise outside the windowed region.

This section begins by introducing conventional LA-based reference-recovery methods~\cite{Li2007a, Segal2010, Niu2018}, which serve as a benchmark for our ML implementation.
We then demonstrate that a complex, ResNet-based autoencoder yields only slightly improved performance at the expense of interpretability, efficiency, and complexity.

%%%%%%%%%%%%%%%%%%%%%%%%%%%%%%%%%%%%%%%%%%%%%%%%%%%%%%%%%%%%%
\subsection{Linear algebra formalism}\label{sec:linear}
%%%%%%%%%%%%%%%%%%%%%%%%%%%%%%%%%%%%%%%%%%%%%%%%
Since the following analysis does not rely on the images' two-dimensional (2D) spatial structure, we work with flattened $P$-dimensional vectors, where $P$ equals the number of pixels.
Vectors and linear operators are denoted in boldface, such as ${\bf v}$ and ${\bf O}$, with components $v_p$ and $O_{pq}$.
The conjugate transpose is denoted by $[{\bf O}^\dagger]_{pq}\equiv O^*_{qp}$ (index reversal with conjugation); and the identity operator is ${\bf I}$.

The basic approach uses a ``training'' set ${\mathcal T} = \{{\bf t}_\alpha\}$ of independently measured reference vectors, collected as the columns of ${\bf T}$, a $P\stimes|\mathcal{T}|$ matrix with elements $T_{p\alpha}$.
In practice, ${\bf T}$ should be sufficiently large that its variability is representative of the underlying distribution.
We adopt the convention that Latin indices ($p$, $q$, $r$, $\dots$) denote image-space (pixel) coordinates, while Greek indices ($\alpha$, $\beta$, $\gamma$, $\dots$) label set elements (training samples, basis vectors, $\dots$).
We then use these vectors to optimally recover each data vector ${\bf d}$ in regions unaffected by the atomic system and extrapolate to the remainder, optimized with respect to a windowed least-squares criterion.

This process requires that the signal of interest be concentrated within an ``interior'' region, isolated by a window ${\bf W}$, a square matrix with eigenvalues $\in[0,1]$ (typically diagonal in the pixel basis, but this need not be the case); the remaining ``exterior'' region is correspondingly isolated by a mask ${\bf M}$ with the same spectral bounds (often ${\bf M} = {\bf I} - {\bf W}$, but this is not required).
For any mask or window ${\bf O}$, we define a metric ${\bf G}_{\bf O} \equiv {\bf O}^\dagger {\bf O}$ (positive-semidefinite Hermitian), which induces the inner product $\langle {\bf u}, {\bf v}\rangle_{\bf G} = {\bf u}^\dagger {\bf G} {\bf v}$ and norm  $||{\bf v}||^2_{\bf G} = \langle {\bf v}, {\bf v}\rangle_{\bf G}$.
In particular, we use the interior, exterior, and Euclidean metrics ${\bf G}_{\bf W}$, ${\bf G}_{\bf M}$, and ${\bf G}_E = {\bf I}$, respectively; furthermore, unannotated inner products and norms (e.g., $\langle {\bf u}, {\bf v}\rangle$) are taken to be Euclidean.

This framing is particularly useful when uncertainties are known in the form of an inverse covariance matrix ${\boldsymbol \Sigma}^{-1}$, which can be expressed as ${\boldsymbol \Sigma}^{-1} = {\bf R}^\dagger {\bf R}$ in terms of the Cholesky factor ${\bf R}$.
For precision-scaled residuals, the following arguments follow through with each metric changed to ${\bf G}_{\bf O} \rightarrow {\bf G}_{{\bf O}{\bf R}} = {\bf R}^\dagger {\bf O}^\dagger {\bf O} {\bf R}$.
In practice, when uncertainties are known, we reconstruct precision-scaled data ${\bf R} {\bf d}$, and use precision-scaled training data.
This removes the need for further discussion of uncertainties and sidesteps the nuance that, while ${\bf W}$ and ${\bf M}$ are fixed, the per-observation uncertainties in general differ.

For every data vector ${\bf d}$, our task is to optimally reconstruct a reference ${\bf r}$ using a basis of linearly independent vectors ${\mathcal B}=\{ {\bf b}_\alpha\}$, collected as the columns of ${\bf B}$, a $P\stimes|{\mathcal B}|$ matrix with elements $B_{p\alpha}$.
Since the basis vectors will be generated from the training set ${\mathcal T}$, we require $|\mathcal B|\le|\mathcal T|$ to avoid underdetermination.
Conventional linear techniques employ a matrix product
\begin{align}
    {\bf r} &= {\bf B}\,{\bf a} ({\bf d})\label{eq:linear:reconstruct}
\end{align}
with amplitudes given by a linear map ${\bf a}({\bf d}) = {\bf A}^\dagger {\bf d}$, defined by the ${P\times|{\mathcal B}|}$ matrix ${\bf A}$.
The elements of ${\mathcal B}$ can further be taken to be orthonormal with respect to a chosen metric, i.e., ${\bf B}^\dagger {\bf G} {\bf B} = {\bf I}$, since any right-factor used to orthonormalize ${\bf B}$ can be absorbed into  ${\bf A}$.

%%%%%%%%%%%%%%%%%%%%%%%%%%%%%%%%%%%%%%%%%%%%%%%%%%%%%%%%%%%%%
\subsection{Basis sets} 
%%%%%%%%%%%%%%%%%%%%%%%%%%%%%%%%%%%%%%%%%%%%%%%%
Both the basis ${\bf B}$ and amplitude map ${\bf A}$ must be determined from the training data ${\mathcal T}$ to best address the question: which set of vectors best describes the interior region (where signal is present) using the amplitudes ${\bf A}^\dagger {\bf d}$ determined from the exterior?
By contrast, the two most common approaches simply reconstruct complete reference images.

The first of these approaches directly uses all $|{\mathcal T}|$ training vectors as the basis, i.e., setting ${\bf B} = {\bf T}$.
By construction, these fully span the training dataset and are operationally linearly independent because $|{\mathcal T}|\lesssim10^3$ is, in practice, much smaller than the dimension $P\approx 10^6$ of the vectors derived from typical $\num{1024}\stimes\num{1024}$ images.

%%%%%%%%%%%%%%%%%%%%%%%%%%%%%%%%%%%%%%%%%%%%%%%%%%%%%%%%%%%%%
\subsection{PCA basis set}\label{sec:linear:pca}
%%%%%%%%%%%%%%%%%%%%%%%%%%%%%%%%%%%%%%%%%%%%%%%%
The second common approach uses principal component analysis (PCA) to generate an orthonormal basis ${\mathcal B}_{\rm PCA}$ with a number of elements $|{\mathcal B}_{\rm PCA}| \leq |{\mathcal T}|$ sufficient to capture the structure present in the training set while suppressing contributions from stochastic noise.
In this simple example, we focus on the case of no mask or window  ${\bf G}_{{\bf M},{\bf W}}={\bf I}$, giving the Euclidean orthonormality condition ${\bf B}^\dagger{\bf B}={\bf I}$.
As shown below, this approach often yields substantial dimensionality reduction, with $|{\mathcal B}_{\rm PCA}| \simeq 16$ even when $|\mathcal T| \simeq \num{1000}$.

In this context, we seek an orthonormal basis of dimension $|{\mathcal B}_{\rm PCA}|$ and transfer matrix ${\bf A}$ that together most effectively describe the training set, with residual vectors ${\boldsymbol \delta} {\bf t}_\alpha \equiv {\bf t}_\alpha - {\bf B} {\bf A}^\dagger {\bf t}_\alpha$.
Collecting these residuals yields the $P\stimes|{\mathcal T}|$ residual matrix ${\boldsymbol \delta} {\bf T} = {\bf T} - {\bf B} {\bf A}^\dagger {\bf T}$.
Given these definitions, the optimal basis minimizes a loss function given by the total variance of the residual vectors
\begin{align}
    L &= \sum_\alpha || {\boldsymbol \delta} {\bf t}_\alpha||^2 = || {\boldsymbol \delta} {\bf T}||^2 = {\rm Tr} \left({\boldsymbol \delta} {\bf T}^\dagger {\boldsymbol \delta} {\bf T}\right)\label{eq:PCA:loss},
\end{align}
where $|| {\bf O} ||$ denotes the Frobenius norm of the matrix ${\bf O}$.
Expanding and using the cyclic property of the trace yields the quadratic loss function
\begin{align}
    L &= {\rm Tr} \left[{\bf T} {\bf T}^\dagger \left({\bf I} - {\bf B} {\bf A}^\dagger - {\bf A} {\bf B}^\dagger  + {\bf A} {\bf A}^\dagger\right)\right],\label{eq:PCA:loss_expanded}
\end{align}
where the first term is the total variance of the training set, and the remainder quantifies the variance explained by the PCA basis.
The required orthonormality constraint adds a Lagrange term ${\rm Tr}\left({\boldsymbol \Lambda}\left({\bf B}^\dagger {\bf B} - {\bf I}\right)\right)$ to $L$, with the Lagrange multiplier matrix ${\boldsymbol \Lambda}$.

Taking matrix derivatives with respect to ${\bf A}^\dagger$ and ${\bf B}^\dagger$ leads to the coupled equations
\begin{align}
    0 &= {\bf T} {\bf T}^\dagger {\bf A} - {\bf T} {\bf T}^\dagger {\bf B} & {\rm and} && 0 &= {\bf B} {\boldsymbol \Lambda} - {\bf T} {\bf T}^\dagger {\bf A} \label{eq:PCA:PCA}
\end{align}
the first is simply solved by setting ${\bf A} = {\bf B}$.
In the second, we identify ${\boldsymbol \Lambda}$ as a diagonal matrix whose entries are the largest $|{\mathcal B_{\rm PCA}}|$ eigenvalues of ${\bf T} {\bf T}^\dagger$ (a positive semidefinite Hermitian matrix with non-negative eigenvalues), and taking the corresponding eigenvectors as the columns of ${\bf B}$.
When every eigenvalue is included, $L\rightarrow 0$, indicating that every training vector can be perfectly recovered.
In practice, this indicates overfitting because each probe image has a nonnegligible contribution from stochastic noise sources such as photon shot noise.

This formulation obscures the fact that, although ${\bf T} {\bf T}^\dagger$ is a $P\stimes P$ matrix, it has at most $|{\mathcal T}|$ nonzero eigenvalues.
We address this by left-multiplying by ${\bf T}^\dagger$ to obtain a new eigenvalue problem
\begin{align}
    0 &= \bar {\bf B} {\boldsymbol \Lambda} - {\bf C} \bar{\bf B},
\end{align}
with the $|{\mathcal T}|\stimes|{\mathcal T}|$ Hermitian matrix ${\bf C} = {\bf T}^\dagger {\bf T}$, a $|{\mathcal T}|\stimes|{\mathcal B}_{\rm PCA}|$ matrix of basis vectors $\bar {\bf B} = {\bf C}^{-1/2} {\bf T}^\dagger {\bf B}$ (with ${\bf C}^{-1/2}$ defined on the support of ${\bf C}$), and the orthonormality condition $\bar {\bf B}^\dagger \bar {\bf B} = {\bf I}$.
This is conveniently solved by the singular value decomposition (SVD) of ${\bf T} = {\bf U} {\boldsymbol \Sigma} {\bf  V}^\dagger$, giving the diagonal singular-value matrix ${\boldsymbol \Sigma} \rightarrow {\boldsymbol \Lambda}^{1/2}$, the right singular vectors ${\bf V}  \rightarrow \bar {\bf B}$, and the left singular vectors ${\bf U}$.  
Thus, the final basis vectors are obtained via the inverse transformation ${\bf B} = {\bf T} {\bf C}^{-1/2} \bar {\bf B} = {\bf U}$.

%%%%%%%%%%%%%%%%%%%%%%%%%%%%%%%%%%%%%%%%%%%%%%%%%%%%%%%%%%%%%
\subsection{Optimal basis sets}\label{sec:linear:optimal}
%%%%%%%%%%%%%%%%%%%%%%%%%%%%%%%%%%%%%%%%%%%%%%%%
We now go beyond the standard PCA case by asking: which exterior basis provides the most information about the interior region, and which basis best describes the interior?
Relative to the PCA calculation in the preceding section, addressing these questions requires two changes.
(i) Including the mask-metric ${\bf G}_{\bf M}$ to the transfer matrix, i.e., giving coefficients ${\bf a} = {\bf A}^\dagger {\bf G}_{\bf M} {\bf d}$ in Eq.~\eqref{eq:linear:reconstruct}.
(ii) Using windowed norms $||{\boldsymbol \delta}{\bf t}_\alpha||_{{\bf G}_{\bf W}}$ in Eq.~\eqref{eq:PCA:loss}.

The PCA solution employed the orthonormality condition ${\bf B}^\dagger{\bf B} = {\bf I}$, which rendered the loss function quadratic in ${\bf B}$, enabling a simple linear-algebra solution.
Here the use of the windowed norm suggests the use of an interior basis matrix ${\bf B}_{\bf W}$ with orthonormality constraints ${\bf B}_{\bf W}^\dagger {\bf G}_{\bf W} {\bf B}_{\bf W} = {\bf I}$.

In the masked-extrapolation setting, we minimize the interior reconstruction loss
\begin{align}
    L &= \sum_\alpha || {\boldsymbol \delta} {\bf t}_\alpha||^2_{{\bf G}_{\bf W}} = {\rm Tr} \left({\boldsymbol \delta} {\bf T}^\dagger {\bf G}_{\bf W} {\boldsymbol \delta} {\bf T}\right)
\label{eq:optimal:loss},
\end{align}
with residuals ${\boldsymbol \delta} {\bf T} = {\bf T} - {\bf B}_{\bf W} {\bf A}^\dagger {\bf G}_{\bf M} {\bf T}$.
Taking matrix derivatives of $L$ with respect to ${\bf A}^\dagger$ and ${\bf B}_{\bf W}^\dagger$ yields the coupled equations
\begin{align*}
0 &= {\bf G}_{\bf M} {\bf T} {\bf T}^\dagger {\bf G}_{\bf W} {\bf B}_{\bf W} -  {\bf G}_{\bf M} {\bf T} {\bf T}^\dagger {\bf G}_{\bf M} {\bf A}\\
0 &=  {\bf G}_{\bf W} {\bf B}_{\bf W} {\boldsymbol \Lambda}_{\bf W} - {\bf G}_{\bf W} {\bf T} {\bf T}^\dagger {\bf G}_{\bf M} {\bf A} 
\end{align*}
where the orthonormality constraint implies Hermitian Lagrange multipliers $ {\boldsymbol \Lambda}_{\bf W}$, giving the required real valued eigenvalues. 

In direct analog with the PCA case, we solve this problem by left-multiplying both expressions by ${\bf T}^\dagger$, obtaining the coupled equations 
\begin{align}
    0 &= {\bf C}_{\bf W}^{1/2} \bar {\bf B}_{\bf W} - {\bf C}_{\bf M}^{1/2} \bar {\bf A}_{\bf M}\\
    0 &=  \bar {\bf B}_{\bf W} {\boldsymbol \Lambda} - \left( {\bf C}^{1/2}_{\bf W} {\bf C}_{\bf M}^{1/2} \right) \bar {\bf A}_{\bf M}
\end{align}
in terms of the $|{\mathcal T}|\stimes|{\mathcal T}|$ Hermitian matrices ${\bf C}_{\bf O} = {\bf T}^\dagger {\bf G}_{\bf O} {\bf T}$, the $|{\mathcal T}|\stimes|{\mathcal B}|$ basis vectors $\bar {\bf B}_{\bf O} = {\bf C}^{-1/2}_{\bf O} {\bf T}^\dagger {\bf G}_{\bf O} {\bf B}_{\bf O}$,  orthonormality constraint $\bar{\bf B}^\dagger_{\bf W} \bar {\bf B}_{\bf W} = {\bf I}$, and with ${\bf O} \in\left\{{\bf W, M}\right\}$.

Again, following the PCA case, the first equation is quickly solved with $\bar {\bf A}_M = {\bf C}_{\bf M}^{-1/2} {\bf C}_{\bf W}^{1/2} \bar {\bf B}_{\bf W}$ leaving behind the eigenvalue equation
\begin{align}
    0 &=  \bar {\bf B}_{\bf W} {\boldsymbol \Lambda} - {\bf C}_{\bf W} \bar {\bf B}_{\bf W}.
\end{align}
This is solved by introducing the interior and exterior SVD expansions of ${\bf O} {\bf T} = {\bf U}_{\bf O} {\boldsymbol \Sigma}_{\bf O} {\bf  V}_{\bf O}^\dagger$, with left singular vectors ${\bf U}_{\bf O}$, right singular vectors ${\bf V}_{\bf O}$ and singular values ${\boldsymbol \Sigma}_{\bf O}$.
All together this gives solutions:
${\boldsymbol \Lambda} = {\boldsymbol \Sigma}_{\bf W}^2$, $\bar {\bf B}_{\bf W} = {\bf V}_{\bf W}$, and $\bar {\bf A}_{\bf M} = {\bf V}_{\bf M} {\boldsymbol \Sigma}_{\bf M}^{-1} {\bf V}^\dagger_{\bf M}  {\bf V}_{\bf W} {\boldsymbol \Sigma}_{\bf W}$.
Or in the initial representation
\begin{align}
    {\bf W} {\bf B}_{\bf W} &= {\bf U}_{\bf W} \\
    {\bf M} {\bf A} &= {\bf U}_{\bf M} {\boldsymbol \Sigma}_{\bf M}^{-1} {\bf V}^\dagger_{\bf M}  {\bf V}_{\bf W} {\boldsymbol \Sigma}_{\bf W}.
\end{align}
If dimension reduction is desired it is implemented by truncating ${\boldsymbol \Sigma}_{\bf W}$ and the columns of ${\bf U}_{\bf W}$ and  ${\bf V}_{\bf W}$.

%%%%%%%%%%%%%%%%%%%%%%%%%%%%%%%%%%%%%%%%%%%%%%%%%%%%%%%%%%%%%
\subsection{Linear algebra implementation}
%%%%%%%%%%%%%%%%%%%%%%%%%%%%%%%%%%%%%%%%%%%%%%%%

%%%%%%%%%%%%%%%%%%%%%%%%%%%%%%%%%%%%%%%%%%%%
\begin{figure*}[t]
\begin{minipage}[t]{0.60\linewidth}
\vspace{0pt}
    \includegraphics{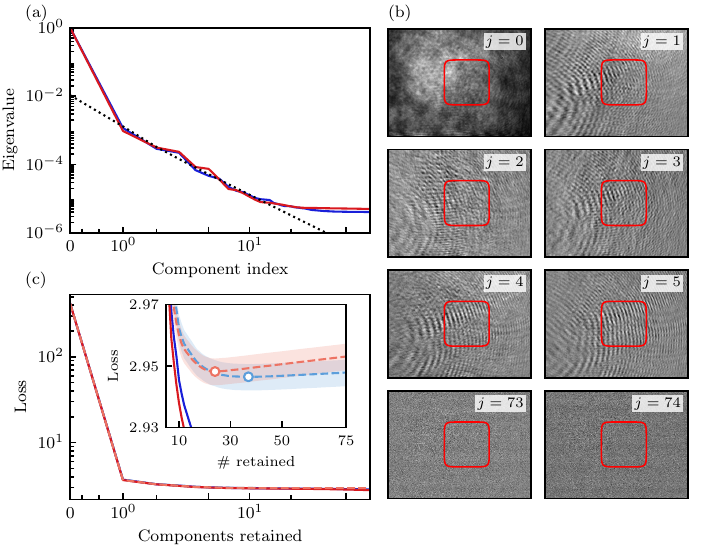}
\end{minipage}
\hfill
\begin{minipage}[t]{0.33\linewidth}
\vspace{-5pt}
    \caption{
    Linear algebra-based reference reconstruction.
    (a) Normalized standard PCA eigenvalues (blue), the optimal linear spectrum $\Sigma_W^2$ (red), and a power-law with exponent $-2$.
    (b) The first 6 and last 2 standard PCA components, with windowed and masked regions delineated by the red squircle.
    (c) Reconstruction loss from 8-fold cross-validation for standard PCA (blue) and the optimal linear (red) methods.
    Training and validation losses are denoted by solid and dashed curves, respectively.
    The inset expands the near-saturation regime and adds $1$-sigma error bands.
    Both validation curves exhibit local minima marked by circles.
    }
    \label{fig:linear}
\end{minipage}
\end{figure*}
%%%%%%%%%%%%%%%%%%%%%%%%%%%%%%%%%%%%%%%%%%%%

We now move from formal derivation to the implementation, performance and interpretability of the two linear methods---``standard PCA'' and ``optimal linear''---before comparing them with more complex deep-learning models.
At the level of numerical linear algebra, both of these algorithms consist of just a few calls to the {\tt BLAS} or {\tt LAPACK} libraries.
While it is true that these libraries do the numerical heavy lifting, considerable effort is required to properly utilize the dataset comprising $\gtrsim 4\times 10^3$ resonant absorption images acquired over about a week of experimental operation.
We therefore begin by exploring the limitations of a simple-minded deployment, then introduce the decomposition of datasets into training, validation, and test sets, and finally introduce $k$-fold cross-validation.

The first results of both reference recovery implementations are shown in Fig.~\ref{fig:linear}a; both standard PCA (blue) and optimal linear (red) each yield a spectrum that, aside from the $j=0$ component, approximately follows a power law with exponent $-2$ (dotted black) before reaching a noise floor\footnote{In our experience, this power-law scaling is ubiquitous, but we are not aware of the underlying mechanism.}. 
Typically, components in the power-law regime contain useful information, while those in the plateau regime are predominantly noise.
This overall impression is confirmed by the first and last few principal components for the standard PCA method shown in Fig.~\ref{fig:linear}b, where the $j=0$ component is essentially the average probe; $j=1$ shows a frequently appearing diffraction ring along with additional features; and $j=2$ includes a phase-shifted diffraction ring with even more artifacts.
By contrast, the final components are dominated by white noise.
This interpretability is possible because the principal components are themselves simple images, i.e., templates, whose importance is given by the spectrum in Fig.~\ref{fig:linear}a.

Each eigenvalue can be interpreted as the reduction in the training loss [Eqs.~\eqref{eq:PCA:loss} or \eqref{eq:optimal:loss}] resulting from that component (this is usually called the explained variance).
Finally, Fig.~\ref{fig:linear}c plots the training loss resulting from the first $j$ components for both the standard PCA (blue) and optimal linear (red) methods\footnote{For conventional PCA, this curve can be obtained by cumulatively summing the eigenvalues.
This does not generalize to the optimal linear method, which is not built from a simple sum over orthogonal components.}
In both cases, the training loss decreases monotonically as more components are included, suggesting that the best strategy is to retain every component.
This is not the case.

In ML language, these examples can be viewed as purely linear models; for example, retaining $75$ components yields a ``model'' with $\approx 12\times10^6$ parameters\footnote{To use standard SVD tooling with our large dataset, the initial $644\times484$ images were down-sampled to $322\times242$.}.
Models of this size are susceptible to overfitting and memorization; in this case, each element of the training dataset contains a non-generalizable component consisting of photon shot noise, which will be learned when a sufficiently large number of components are retained.

%%%%%%%%%%%%%%%%%%%%%%%%%%%%%%%%%%%%%%%%%%%%
\begin{figure*}[t]
\raggedright
\begin{minipage}[t]{0.33\linewidth}
\vspace{-5pt}
    \caption{
    ML model modalities for reference recovery.
    (a) A simple encoder-decoder.  
    When no nonlinear activation functions are present, this reduces to the linear models introduced in Secs.~\ref{sec:linear:pca} and \ref{sec:linear:optimal}. 
    (b) A U-Net with shortcut connections.
    In both panels, rectangles denote data vectors and trapezoids denote layers, oriented to indicate the typical direction of dimension reduction or expansion.
    }    \label{fig:models}
\end{minipage} 
\hspace{0.01\linewidth}
\begin{minipage}[t]{0.60\linewidth} 
\vspace{0pt}
    \includegraphics{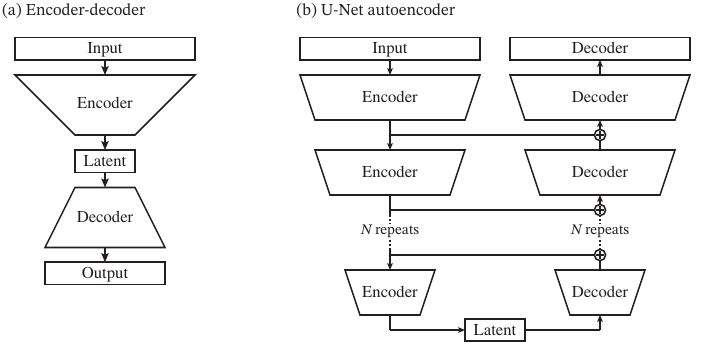}
\end{minipage}
\end{figure*}
%%%%%%%%%%%%%%%%%%%%%%%%%%%%%%%%%%%%%%%%%%%%

Data science resolves these types of issues by first randomly splitting the overall dataset into $k\approx10$ subsets, designating one subset for validation, and leaving the remainder for training.
When repeated for all $k$ possible validation sets, this is called $k$-fold cross-validation.
The dashed curves in Fig.~\ref{fig:linear}c (only visible in the inset) show 8-fold cross-validation losses for both the standard (blue) and optimal (red) methods along with error bands reflecting the variation across the folds. 
Unlike the training loss, which monotonically decreases, the validation loss in the inset first saturates, and then increases: this is typical over-training behavior.
This analysis identifies the optimal number of components to retain, $37$ for the standard PCA method and $24$ for the optimal linear method.
In this case, the added sophistication of the optimal-linear approach pays off, not in improved ultimate performance, but rather a reduction in overall parameter count.  
These linear models serve as baselines for the non-linear models introduced next.

In a more complete ML workflow, one would first set aside a randomly selected subset, $\approx 10~\%$ of the full dataset, as a test set that is not used during model design, training, or validation.
For demonstration purposes in this chapter, we did not set aside a test set.

%%%%%%%%%%%%%%%%%%%%%%%%%%%%%%%%%%%%%%%%%%%%%%%%%%%%%%%%%%%%%
\subsection{ML model modalities}
%%%%%%%%%%%%%%%%%%%%%%%%%%%%%%%%%%%%%%%%%%%%%%%%
From an ML viewpoint, both of our linear methods are minimal encoder-decoder models with two dense layers (one input and one output), no activation functions, and with equal input and output dimensions; in other words, they are purely linear autoencoders, schematically shown in Fig.~\ref{fig:models}a.
In abstract terms, these models linearly map a $P$-dimensional input vector ${\bf d}$ (flattened image) to a lower-dimensional latent vector ${\bf a}$ of dimension $|\mathcal{B}|$, and then employ a second linear transform to produce a reconstructed $P$-dimensional output vector [e.g., Eq.~\eqref{eq:linear:reconstruct}].

The overall encoder-decoder architecture is agnostic to the specific contents of the ``encoder'' and ``decoder blocks.''
For example, using the notation introduced in Sec.~\ref{sec:linear}, one such block could be a standard dense layer
\begin{align}
    a_p &= \phi([{\bf A} {\bf d} + {\bf b}]_p) = f\left(\sum_q A_{pq} d_q + b_p\right)\label{eq:dense}
\end{align}
which combines a linear transformation ${\bf A}$, an additive bias ${\bf b}$, and a pointwise nonlinear {\it activation function} $\phi$ applied to each output (see Sect.~\ref{sec:cnn} for common choices).
A single such layer is called a perceptron, and stacking two or more perceptrons gives the multilayer perceptron (MLP) architecture~\cite{Goodfellow2016}, one of the canonical deep-neural-network architectures.

The first stage of these models flattens 2D images into simple vectors, discarding implicit spatial structure and forcing the model to relearn any relevant spatial relations.
This is inefficient, often requiring a large number of parameters and long training times when direct methods such as SVD are unavailable.
CNNs overcome both limitations by employing small convolutional kernels that operate directly in image space.
This replaces the ``encoder'' stage of an encoder-decoder with one or more convolutional layers.
Likewise, the ``decoder'' stage can be replaced by convolutional-like layers such as ConvTranspose or PixelShuffle.
ConvTranspose is an approximate deconvolution layer, and PixelShuffle operates by first using a CNN to grow its 1D depth-wise channels and then reshaping them into a 2D image.

Published ML-based solutions to the reference recovery task typically employ the more sophisticated U-Net architecture shown in Fig.~\ref{fig:models}b.
Originally developed for image segmentation, the U-Net extends the autoencoder pattern with shortcut connections and often gives high-quality results even with modest-sized training sets~\cite{Ronneberger2015}.
While the U-Net pattern can, in principle, use generic encoder and decoder blocks, it typically uses CNNs for encoding and ConvTranspose or PixelShuffle layers for decoding.

%%%%%%%%%%%%%%%%%%%%%%%%%%%%%%%%%%%%%%%%%%%%%%%%%%%%%%%%%%%%%
\subsection{ML implementation}
%%%%%%%%%%%%%%%%%%%%%%%%%%%%%%%%%%%%%%%%%%%%%%%%
With this backdrop, we finally turn to an ML implementation of the reference recovery problem.
For brevity, this demonstration augments the encoder-decoder pattern~\footnote{As opposed to the U-Net with shortcut connections, this structure is better able to filter noise processes such as shot noise and readout noise.} with a latent space ResNet~\cite{He2016} to deliver a modest performance improvement.
This comes at the expense of drastically increased implementation overhead, including model design, regularization, data augmentation, hyperparameter selection, and training.

%%%%%%%%%%%%%%%%%%%%%%%%%%%%%%%%%%%%%%%%%%%%
\begin{figure}[t]
    \includegraphics{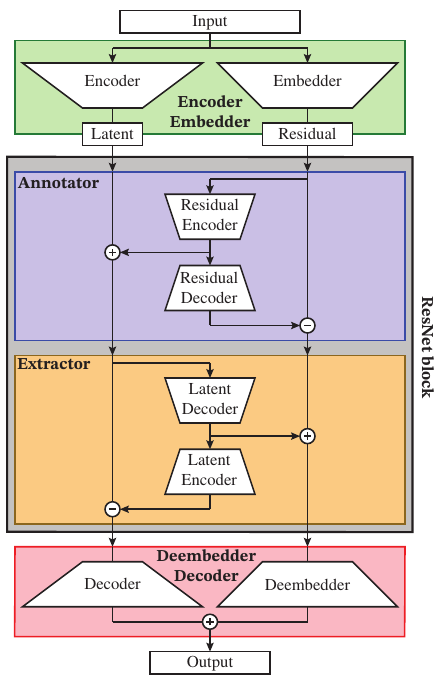}
    \caption{
    ResNet encoder-decoder. 
    Data-flow is indicated by lines, while dimension-changing layers are denoted by trapezoids.
    Functional groups are denoted by larger rectangles.
    }
    \label{fig:models:resnet}
\end{figure}
%%%%%%%%%%%%%%%%%%%%%%%%%%%%%%%%%%%%%%%%%%%%

\begin{trailer}{Model design}
Several model architectures were explored before converging on the ResNet encoder-decoder shown in Fig.~\ref{fig:models:resnet}.
The ResNet encoder-decoder model organizes individual encoders and decoders (trapezoids) into high-level functional blocks with two distinct low-dimensional data pathways.
The ``latent'' path uses purely linear encoders and decoders (in the encoder-embedder and deembedder-decoder stages); in isolation, it recovers the performance of the linear models.
The ``residual'' path is transformed to and from image space by single-layer perceptrons (in the encoder-embedder and deembedder-decoder stages).
ResNet blocks then exchange information between these two data paths, where the latent path carries the already highly predictive linear reference-recovery information, while the residual path corrects for nonlinear artifacts left behind by the linear model.

At the architecture's level, this model introduces just three hyperparameters: the number of ResNet blocks {\tt NUM\_RESNET}, and the dimensions of the data paths {\tt DIM\_LATENT} and {\tt DIM\_RESIDUAL}.
\end{trailer}

\begin{trailer}{Regularization and data-augmentation}
Larger or more complicated models are often prone to over-training and generally require large datasets.
In this context, regularization methods combat overtraining and accelerate training, while data augmentation expands the effective size of the dataset (which can also reduce overtraining). 

Our example ResNet encoder-decoder model-level implements regularization with both a Dropout layer (prior to the encoder-embedder block) and BatchNorm layers (following the residual encoder and residual decoder layers).
A Dropout layer randomly zeros elements of the data vector between layers (and rescales them to maintain the overall signal level).
BatchNorm layers rescale data vectors so that each component has a unit standard deviation across the training dataset.
The optimal location of these layers in the model is determined empirically.
Data augmentation can be quite sophisticated, for example, generating data with auxiliary models.
For simplicity, we augment by including small spatial displacements of the images prior to flattening.

This introduces two more hyperparameters: the dropout fraction {\tt DROPOUT\_FRAC}, and the maximum augmentation shift {\tt MAX\_SHIFT}.
\end{trailer}

\begin{trailer}{Training}
In contrast to the linear models, the addition of nonlinearities and hidden layers makes the loss landscape complex, with many local minima.
This is a generic problem in deep learning, and has spawned many gradient-based optimizers, with well-known examples including Adam, AdamW, and Lion.
These optimizers themselves contain hyperparameters, such as the learning rate, and, in the case of AdamW, additional regularization parameters.
AdamW was found to outperform other options for this task; the weight decay present in AdamW controls the growth of large eigenvalues in the linear Encoder/Decoder stages of our ResNet model.

In addition, training follows a schedule in which one or more parameters are adjusted throughout training.
For example, here we begin with a OneCycle stage for \num{1000} epochs that quickly increases and decreases {\tt ETA}, followed by a linear decrease for \num{9000} epochs, and concluding with a power-law with exponent $-1$ for the remainder of training.

All together, this introduces several more hyperparameters: the overall learning rate {\tt ETA} and the weight decay {\tt LAMBDA}, in addition to the many parameters describing the training schedule.
All hyperparameters were found empirically.
\end{trailer}

\begin{trailer}{Hyperparameter selection}
All hyperparameters must be individually tuned for optimal training and final model performance.
This requires a second stage of optimization wrapped around the minimization problem solved by training.  
In practice, one finds coarse values for these using individual training runs, optimizing on the validation loss, and then switches to longer $k$-fold cross-validation runs to identify the final values.

For example, {\tt DIM\_LATENT} and {\tt DIM\_RESIDUAL} were initially set to the ideal number of components retained in the linear models, and then optimal performance was found by searching away from this point.
Searching across the full hyperparameter space can be time-consuming; in this case, it requires testing 100+ hyperparameter combinations to achieve acceptable performance and reasonable training speed.
\end{trailer}

%%%%%%%%%%%%%%%%%%%%%%%%%%%%%%%%%%%%%%%%%%%%
\begin{figure}[t]
    \includegraphics{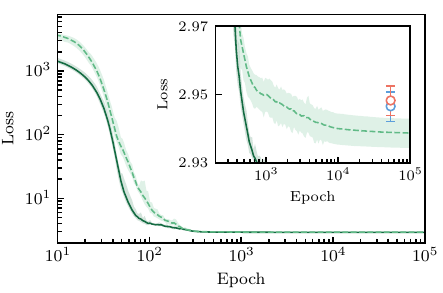}
    \caption{
    ResNet training with 8-fold cross-validation. 
    Training and validation data are denoted by solid and dashed curves, respectively, and include $1$-sigma error bands.
    The inset zooms in on the fine-tuning stage of training, and the markers show the results for the linear models in Fig.~\ref{fig:linear} for conventional PCA (blue) and optimal linear (red).
    }
    \label{fig:resnet}
\end{figure}
%%%%%%%%%%%%%%%%%%%%%%%%%%%%%%%%%%%%%%%%%%%%

The $8$-fold cross-validation training curves for the final ResNet encoder-decoder model are shown in Fig.~\ref{fig:resnet}, with training and validation data indicated by the solid and dashed curves, respectively.
In both cases, the loss drops by orders of magnitude in the first $\approx 100$ epochs before slowing down; by epoch \num{1000}, training has drastically slowed, entering into a fine-tuning phase.
As seen in the inset, the validation loss continues to decrease after $10^5$ epochs and is slightly lower than that of the linear models (markers); the final training loss is well below the validation loss, indicating some degree of memorization.
The ResNet uses the same reconstruction loss [Eq.~\eqref{eq:optimal:loss}], preprocessing, image resolution, and cross-validation folds as the linear benchmarks, so differences in Figs.~\ref{fig:linear} and \ref{fig:resnet} are attributable to the model rather than the data pipeline.

Careful inspection of the images in Fig.~\ref{fig:imaging}c confirms that the background is slightly less noisy for the ML case.
Still, this improvement came at the expense of vastly increased effort (days rather than weeks), both at the human and computational levels.
Furthermore, the limited explainability of the linear models has been sacrificed.

The main conclusion is that ML methods can offer improvements upon standard numerical methods, but the gains can be modest, and the final value proposition should be considered. 
For example, it is much more straightforward to integrate LA-based reference recovery into daily laboratory operations, where real-time data analysis informs decision-making.
Further, experiments focusing on the overall density distribution, such as the Gaussian thermal distributions in Fig.~\ref{fig:imaging}c, would seldom benefit from the often negligible improvements offered by ML models.
However, experiments studying the pixel-by-pixel fluctuations of the atom signal away from its mean (i.e., atom shot noise) are an ideal use case for ML tools, as they often operate in regimes where the imaging noise scale exceeds the signal of interest.

%%%%%%%%%%%%%%%%%%%%%%%%%%%%%%%%%%%%%%%%%%%%%%%%%%%%%%%%%%%%%
\section{Learning with understanding: Case of solitonic excitations}\label{sec:solitons}
%%%%%%%%%%%%%%%%%%%%%%%%%%%%%%%%%%%%%%%%%%%%%%%%%%%%%%%%%%%%%
Solitons are robust, localized wave structures that, despite waves' typical tendency to spread, preserve their shape as they move through their host medium.
They appear in many nonlinear media, such as water~\cite{Russel1837, Lakshmanan2009}, optical fibers~\cite{Hasegawa2000}, and plasmas~\cite{Kono2010}.
Bose-Einstein condensates offer an unusually simple, clean, and controllable environment in which to study solitonic excitations.
This is because over a wide range of conditions, the many-body system can be described by a single macroscopic wavefunction with a well-defined amplitude and phase~\cite{Dalfovo1999}.
The amplitude encodes the atomic density, while spatial variations of the phase describe superfluid flow. 
In this language, a solitonic excitation is not merely a \textit{dip} or a \textit{bump} in density; it is a coupled structure in both density and phase, stabilized by a balance between dispersion (which tends to broaden a wave packet) and nonlinear interactions (which can self-steepen or self-focus the wave).

%%%%%%%%%%%%%%%%%%%%%%%%%%%%%%%%%%%%%%%%%%%%
\begin{figure*}[t]
    \centering
    \includegraphics{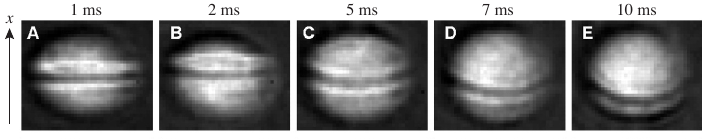}
    \caption{
    Images showing evolution of the BEC (measured after TOF) from 1~\si{\milli\second} (panel A) to 10~\si{\milli\second} (panel E) after imprinting a phase step of about $1.5\pi$ on the top half of the condensate. 
    A positive density disturbance is moving rapidly in the $+x$ direction, while a dark solitonic excitation (most likely a vortex ring) is moving significantly slower in the $-x$ direction (opposite to the direction of the applied force).
    Reproduced with permission from Ref.~\cite{Denschlag2000}.
    }
    \label{fig:kink-examples}
\end{figure*}
%%%%%%%%%%%%%%%%%%%%%%%%%%%%%%%%%%%%%%%%%%%%

Repulsively interacting 1D (or quasi-1D) BECs support \textit{dark} (or \textit{gray}) solitons, of which the static black soliton is a special case.
They can be thought of as moving ``phase defects'' that produce a localized depletion in the atomic density~\cite{Frantzeskakis2010}.
A black soliton is a static solution that fully depletes the condensate density and is accompanied by a full $\pi$ phase jump.
Gray solitons are moving solutions of increased width that only partly deplete the density, and have a phase jump that drops to zero (or increases to $2\pi$) as their velocity increases to the condensate speed of sound~\cite{Busch2000}.
Seminal experiments demonstrated that such solitons can be created in a controlled way---for example, by imprinting a spatial phase pattern onto a BEC (``phase imprinting'' or ``phase engineering'') and then watching the resulting excitation propagate~\cite{Denschlag2000}.
As noted in the introduction, attractively interacting BECs have bright, rather than dark, soliton solutions.

In elongated (cigar-shaped) condensates---where the axial dynamics are often much slower than the tightly confined radial dynamics---these excitations often appear as narrow, stripe-like density depletions in TOF imaging (see Sec.~\ref{sec:imaging}).
As we mentioned in the \textit{Introduction}, several practical challenges complicate the analysis of such images: the observed contrast and apparent width depend on the soliton's velocity and on the experimental imaging resolution; thermally excited atoms can obscure the signal, while interference fringes add distortions; and there is no single reliable visual template for identifying solitonic excitations.
Moreover, the acquired 2D images are typically projections of a 3D cloud, meaning that a localized 3D structure can appear as a deceptively simple 2D stripe.
As a result, the same physical excitation can look different across experimental conditions, and distinct excitations can look superficially similar in a single-shot image.

For high-throughput analysis, it is often useful to adopt labels that reflect what is reliably distinguishable in the imaging modality at hand.
In TOF images of elongated BECs, one can define coarse classes based on the presence or absence of well-defined features as ``no excitation,'' ``single excitation,'' and ``other excitations'' (e.g., multiple stripes, low-contrast events, or otherwise ambiguous structures)~\cite{Guo2021, Fritsch2022}.
Within the single-excitation category, using additional data-processing approaches, such as modeling and physics-informed fitting, allows for introducing finer distinctions between longitudinal solitons\footnote{This class includes any feature whose projection is a clean density depletion, potentially including: kink solitons, longitudinally aligned solitonic vortices, and large-radius vortex-rings to name a few.} and (transverse) solitonic vortices~\cite{Guo2022}.
In the following section, we briefly overview the most common types of observed excitations in elongated BECs.

%%%%%%%%%%%%%%%%%%%%%%%%%%%%%%%%%%%%%%%%%%%%%%%%%%%%%%%%%%%%%
\subsection{A short guide to soliton types in elongated BEC experiments}
%%%%%%%%%%%%%%%%%%%%%%%%%%%%%%%%%%%%%%%%%%%%%%%%
Since many experiments with BECs are conducted in highly elongated traps, it is natural to first model the excitation as effectively one-dimensional motion along the long axis.
However, real condensates are only \textit{quasi}-1D rather than strictly 1D, and the residual three-dimensional structure can matter qualitatively.
Planar dark solitons in three dimensions can be dynamically unstable and can evolve into other long-lived solitary-wave structures~\cite{Anderson2001}.
A useful way to organize the landscape is therefore to distinguish solitons that are well approximated as quasi-1D kinks in density from 3D solitary waves that retain soliton-like density signatures but carry vorticity and more complex geometry, such as solitonic vortices, vortex rings, and more complex structures~\cite{Anderson2001, Donadello2014}.

%%%%%%%%%%%%%%%%%%%%%%%%%%%%%%%%%%%%%%%%%%%%%%%%
\noindent
{\bf Dark and gray kink solitons}---These are canonical ``soliton'' excitations: a localized depletion accompanied by a change in the phase that becomes a step for a static black soliton
In images of elongated BECs, they often appear as a stripe-like depletion roughly transverse to the long axis, see Fig.~\ref{fig:kink-examples}.
Slow solitons tend to appear deeper and more localized, while faster solitons are typically shallower and harder to distinguish from background structure.
In our experimental pipeline, the term \textit{kink soliton} is used to emphasize this phase-change character, even when the imaging signature is simply a density notch~\cite{Aycock2017, Frantzeskakis2010}.
More broadly, once three-dimensional effects become relevant, it is best to think in terms of a family of related solitary-wave excitations rather than a single ``soliton'' archetype~\cite{Mateo2015}.

\begin{figure*}[t]
\centering
\begin{minipage}[t]{0.60\linewidth}
\vspace{0pt}
    \includegraphics{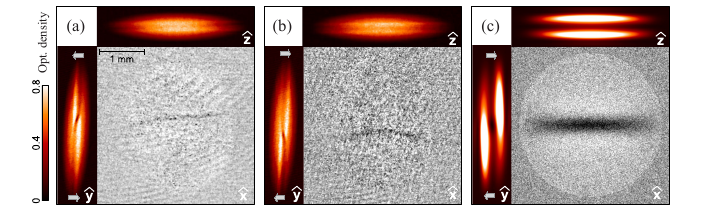}
\end{minipage}
\hfill
\begin{minipage}[t]{0.33\linewidth}
\vspace{-5pt}
    \caption{
    (a)--(b) Experimental images of the integrated density distribution of two BECs after TOF of $120$~\si{\milli\second}, revealing the presence of a solitonic vortex with opposite circulation. 
    (c) Theoretical 3D calculation for the experimental configuration, with clockwise circulation.
    In all panels, arrows indicate the atomic flow.
    Reproduced with permission from Ref.~\cite{Donadello2014}.
    }
    \label{fig:vortex-examples}
\end{minipage}
\end{figure*}

%%%%%%%%%%%%%%%%%%%%%%%%%%%%%%%%%%%%%%%%%%%%%%%%
\noindent
{\bf Solitonic vortices}---A solitonic vortex is a vortex line embedded in an elongated BEC whose density depletion can resemble a soliton sheet, but whose phase twists around a core due to the phase gradient. 
Experimentally, solitonic vortices can leave distinctive signatures after expansion, such as a twisted planar density depletion, and interferometric measurements can reveal dislocations associated with circulation~\cite{Donadello2014}.
For image-based classification, the important lesson is that a stripe-like depletion does not uniquely identify a kink soliton: vortical excitations can masquerade as solitons in projection, see Fig.~\ref{fig:vortex-examples}.

%%%%%%%%%%%%%%%%%%%%%%%%%%%%%%%%%%%%%%%%%%%%%%%%
\noindent
{\bf Vortex rings and other 3D solitary waves}---Vortex rings are closed-loop vortex lines that can also appear as localized depletions, sometimes evolving from unstable planar solitons~\cite{Anderson2001}.
Depending on geometry and imaging axis, a vortex ring may appear as a pair of depletions, a partially filled notch, or a more complex structure.

To summarize, solitonic excitations are rich but visually heterogeneous: their appearance varies with velocity, dimensionality, and imaging artifacts, and different excitations can project to similar stripe-like patterns.
This is exactly the setting where one might want to maintain the balance between high-throughput classification and localization---CNNs excel here---and model decisions that can be audited against physics---which is where interpretable models become attractive.
In the following sections, we discuss these two settings in more depth.
We also introduce the simplified dark solitons dataset we use for all demonstrations.

%%%%%%%%%%%%%%%%%%%%%%%%%%%%%%%%%%%%%%%%%%%%%%%%%%%%%%%%%%%%%
\subsection{The dark solitons in BECs dataset 2.0}\label{sec:soliton_dataset}
%%%%%%%%%%%%%%%%%%%%%%%%%%%%%%%%%%%%%%%%%%%%%%%%
The dataset used for experiments discussed later in this chapter is a subset of the \textit{Dark solitons in BECs dataset} (v2.0)~\cite{SolitonDataset}.
This dataset was originally created to support the development and benchmarking of ML methods for cold-atom experiments and to provide a concrete testbed at the interface of ML and quantum physics~\cite{Fritsch2022}.
It consists of more than $1.6\times 10^4$ pre-processed experimental images of BECs after TOF, acquired under a range of conditions, some with and some without solitonic excitations present.
Since the images are produced by a real experimental apparatus, they naturally include the variability and imperfections that make automated analysis challenging and scientifically relevant: fluctuations in atom number and cloud shape, imaging noise, and genuine shot-to-shot differences in the excitations themselves.

%%%%%%%%%%%%%%%%%%%%%%%%%%%%%%%%%%%%%%%%%%%%
%%%%%%%%%%%%%%%%%%%%%%%%%%%%%%%%%%%%%%%%%%%%

\begin{figure*}[t]
\raggedright
\begin{minipage}[t]{0.33\linewidth}
\vspace{-5pt}
    \caption{
    (a)--(c) Raw data showing (a) probe with atoms $I$, (b) probe only $I_{\rm p}$, and (c) background $I_{\rm bg}$ (scaled by $100\times$).
    As in Fig.~\ref{fig:imaging}, red boxes outline the region where the BEC is located.
    This data was used to derive the optical depth (e).
    (d)--(f) Representative preprocessed images from the \textit{Dark solitons in BECs dataset 2.0} showing a BEC (d) without excitations (\texttt{class-0}), (e) with a lone excitation (\texttt{class-1}), and 
    (f) with multiple excitations (\texttt{class-2}).
    The green arrows indicate all solitonic excitation positions.
    (g)--(i) Overall column-integrated optical depth (gray) and corresponding fluctuations away from the nominal Thomas-Fermi profile (black) for a BEC (g) without excitations, (h) with a lone excitation, and (i) with multiple excitations.
    The green dashed lines mark the location of the depletions in the density fluctuations corresponding to the solitonic excitation indicated in (d)--(f).
    }
    \label{fig:data_sample}
\end{minipage}
\begin{minipage}[t]{0.60\linewidth}
\vspace{0pt}
    \includegraphics{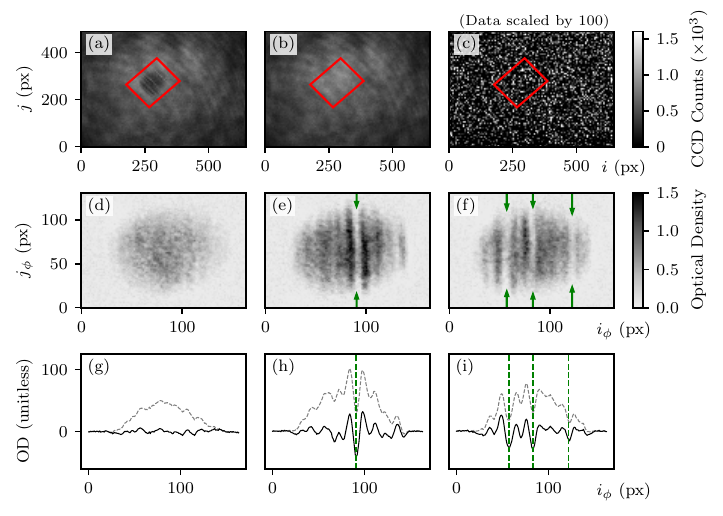}
\end{minipage}
\end{figure*}

As introduced in Sec.~\ref{sec:imaging}, these data were obtained via absorption imaging after TOF\footnote{After accounting for the magnification of our imaging system, the $648\times488$ pixel raw images (Point Grey FL3, $5.6$~\si{\micro\meter} pixel pitch) have an effective pixel size of $0.93$~\si{\micro\meter}, appreciably smaller than our $2.8$~\si{\micro\meter} optical resolution.}, yielding the raw data as in Figs.~\ref{fig:data_sample}(a)--(c).
The condensate, visible in the red rectangle in Fig.\ref{fig:data_sample}a,
occupies only a small fraction of the field of view.
Figures~\ref{fig:data_sample}(d)--(f) show the optical depth computed using Eq.~\eqref{eq:OD} after rotating by $\phi \approx 45^\circ$ to align with the long axis of the BEC, with rotated coordinates denoted by $[i_\phi, j_\phi]$.
To produce standardized inputs suitable for both manual labeling and ML training, each optical density image is then fit to a column-integrated Thomas-Fermi profile
\begin{widetext}
\begin{align}\label{eq:2Dfit}
    n^{\rm TF}_{i,j} = n_{0}\,{\max}\left\{\left[1-\left(\frac{i_\phi-i_0}{R_i}\right)^2-\left(\frac{j_\phi-j_0}{R_j}\right)^2\right],0\right\}^{3/2} + \delta n,
\end{align}
\end{widetext}
providing estimates of the cloud center $[i_0, j_0]$, the peak 2D density $n_0$, the Thomas--Fermi radii $[R_i, R_j]$, along with an overall offset $\delta n$ to correct for the small changes in probe intensity between images.
The images are cropped to a fixed size of $164\times132$ pixels centered at $[i_0, j_0]$.
Finally, an elliptical mask (with semi-major and minor radii determined from the Thomas-Fermi fit) is applied to suppress background noise outside the cloud.
Together, these steps enforce consistent geometry across shots, reduce irrelevant pixel variation, and make the remaining intensity structure more directly comparable across experimental conditions.

%%%%%%%%%%%%%%%%%%%%%%%%%%%%%%%%%%%%%%%%%%%%%%%%
\noindent
{\bf Coarse labels: presence of excitations}---The \textit{Dark solitons in BECs dataset} (v2.0) provides several types of labels to support training ML models for automated analysis.
The first category of labels is a deliberately simple set of human assigned coarse labels that capture the most robust visual distinctions available~\cite{Guo2021}: \texttt{class-0} (Fig.~\ref{fig:data_sample}d), indicating no excitations; \texttt{class-1} (Fig.~\ref{fig:data_sample}e), indicating a single, well-defined excitation; and \texttt{class-2} (Fig.~\ref{fig:data_sample}f), indicating shots that are neither unambiguously empty nor clean single-excitation cases (e.g., multiple stripes, additional density modulations, or ambiguous structures).
This three-way split mirrors the practical decision an experimentalist typically makes when scanning large image sets and provides a stable foundation for both benchmarking and downstream, more detailed labeling.
During data curation, an additional class was introduced, \texttt{class-8}, to indicate data flagged during the process as potentially mislabeled~\cite{Fritsch2022}.
Roughly $60~\%$ of the v2.0 is unlabeled and assigned to \texttt{class-9}.

%%%%%%%%%%%%%%%%%%%%%%%%%%%%%%%%%%%%%%%%%%%%
%%%%%%%%%%%%%%%%%%%%%%%%%%%%%%%%%%%%%%%%%%%%

%%%%%%%%%%%%%%%%%%%%%%%%%%%%%%%%%%%%%%%%%%%%%%%%
\noindent
{\bf Fine labels: physics-informed excitation categories}---The \texttt{class-1} data, i.e., with exactly one identified excitation, are annotated with a finer, physically motivated taxonomy stored as the \texttt{excitation\_PIE} label~\cite{Guo2022}.
This additional label is provided by a physics-informed excitation (PIE) classifier, which uses parameters of a Ricker wavelet (i.e., the Mexican hat wavelet) fitted to the residual density (condensate density profile with Thomas-Fermi fit subtracted) to assign one of six excitation categories within \texttt{class-1}.

\begin{figure*}[t]
\begin{minipage}[t]{0.60\linewidth}
\vspace{0pt}
    \includegraphics{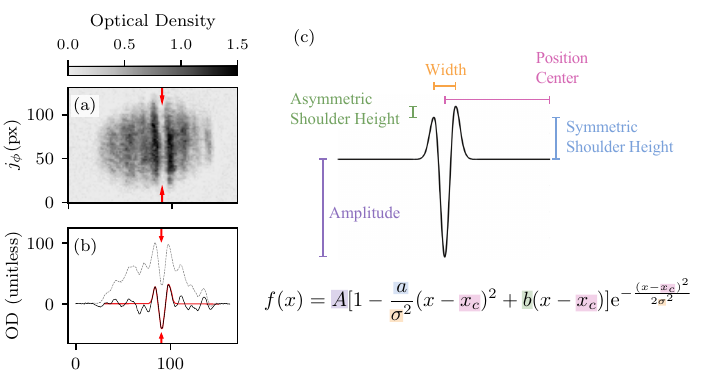}
\end{minipage}
\hfill
\begin{minipage}[t]{0.33\linewidth}
\vspace{-5pt}
    \caption{
    Physics-informed feature extraction for solitonic excitations.
    (a) Image of a BEC with a solitonic excitation identified by the red arrows.
    (b) 1D density profiles, both with (black) and without (grey) Thomas-Fermi background subtraction.
    The red curve plots a fitted Mexican-hat model capturing the localized depletion structure.
    (c) Schematic of the fit-model illustrating the key five parameters: amplitude $A$, center $x_c$, width $\sigma$, and symmetric and asymmetric shoulder heights $a$ and $b$.
    These parameters provide a compact, physically interpretable description of the excitation and form the basis of the feature representation used for classification.
    }
    \label{fig:pie_params}
\end{minipage}
\end{figure*}

\begin{figure*}[t]
\raggedright
\begin{minipage}[t]{0.33\linewidth}
\vspace{-5pt}
    \caption{
    Example images of data from the six PIE classes: (a) a longitudinal soliton, (b) canted excitation, (c) top partial excitation, (d) bottom partial excitation, (e) clockwise vortex, and (f) counterclockwise vortex.
    The vorticity in (e) and (f) is assigned based on how the density fronts curve around the excitation axis.
    The green arrows indicate all excitation positions.
    }
    \label{fig:pie_classes}
\end{minipage}
\begin{minipage}[t]{0.60\linewidth}
\vspace{0pt}
    \includegraphics{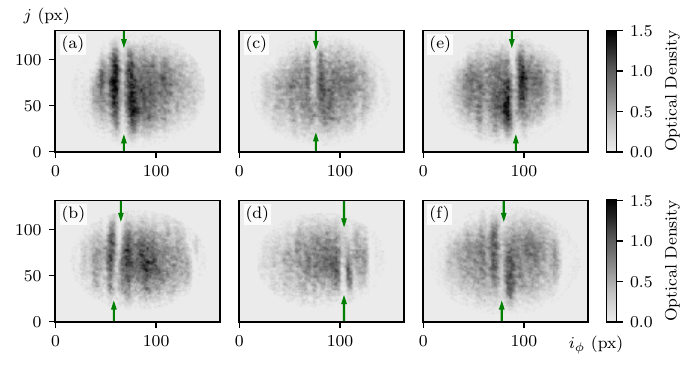}
\end{minipage}
\end{figure*}

The physics-informed feature representation used throughout this section is illustrated in Fig.~\ref{fig:pie_params}.
Starting from the OD image of a BEC containing a lone solitonic excitation (Fig.~\ref{fig:pie_params}a), we extract a 1D density profile along the analysis axis and subtract the Thomas-Fermi background to isolate the localized depletion (Fig.~\ref{fig:pie_params}b).
This residual profile is then fit with a simplified Mexican-hat model (Fig.~\ref{fig:pie_params}b, red).
The model's parameters provide a compact and physically meaningful description of the excitation, see Fig.~\ref{fig:pie_params}c.
In particular, the amplitude $A$, width $\sigma$, position $x_c$, and symmetric $a$ and $b$ asymmetric shoulder parameters derived independently from the top and bottom half of the image capture the key geometric and structural properties that distinguish different excitation types.
The ten Mexican hat fit parameters are used by the PIE classifier.
Figure~\ref{fig:pie_classes} shows examples of BECs from all six PIE classes.

%%%%%%%%%%%%%%%%%%%%%%%%%%%%%%%%%%%%%%%%%%%%%%%%%%%%%%%%%%%%%
\subsection{Problem setup}\label{sec:problem_setup}
%%%%%%%%%%%%%%%%%%%%%%%%%%%%%%%%%%%%%%%%%%%%%%%%
In the remainder of this chapter, we focus on the \texttt{class-1} subset of the dataset.
The fine-grained \texttt{excitation\_PIE} labels are strongly imbalanced, with the dominant \textit{longitudinal soliton} category containing 2,229 images and the smallest \textit{clockwise vortices} category containing only 28 images.
Thus, we apply several curation steps common in data science to reduce the extreme imbalance while preserving the physics.

First, we remove images labeled as \textit{canted}, as a combination of orientation variability and small class size makes the category difficult to learn reliably.
Second, we merge physically symmetric categories into consolidated classes, i.e., we combine \textit{top} and \textit{bottom} partial soliton classes into a single \textit{partial} category, and \textit{clockwise} and \textit{counterclockwise} solitonic vortices into a single \textit{vortex} category.
To ensure consistency within the merged classes, the \textit{bottom partial} and \textit{counterclockwise vortex} images are transformed by symmetry operations so that their visual appearance is always \textit{top partial} and \textit{clockwise vortex}, respectively (i.e., the dataset contains no bottom partial or counterclockwise vortex examples).
We refer to the resulting three-class subset as the \textit{reduced soliton dataset}.
The reduced set provides a realistic, experimentally grounded benchmark for comparing interpretable EBM pipelines trained on low-dimensional, physics-informed image representations with image-native CNN models.
Our goal is to train a classifier that maps the observed data to a probability distribution over labels while remaining robust to common experimental imperfections, such as shot-to-shot variations in atom number, small translations, and residual imaging artifacts.

%%%%%%%%%%%%%%%%%%%%%%%%%%%%%%%%%%%%%%%%%%%%
%%%%%%%%%%%%%%%%%%%%%%%%%%%%%%%%%%%%%%%%%%%%

%%%%%%%%%%%%%%%%%%%%%%%%%%%%%%%%%%%%%%%%%%%%%%%%%%%%%%%%%%%%%
\subsection{A brief introduction to Explainable Boosting Machines}
%%%%%%%%%%%%%%%%%%%%%%%%%%%%%%%%%%%%%%%%%%%%%%%%
PCA enforces $\ell_2$-orthogonality among the principal components, making it a powerful tool for compactly representing high-dimensional images and capturing dominant modes of variance.
However, orthogonality in pixel space does not necessarily translate to semantic complementarity of the underlying physical structures.
Even the leading components might mix multiple sources of variation, such as condensate shape fluctuations and diffraction or fringe patterns, as seen in the first six principal-component images in Fig.~\ref{fig:linear}b.
This mixing complicates interpretation. 
Thus, while PCA provides an efficient coordinate system, it does not by itself yield a transparent explanation of which physically meaningful cues drive a downstream classification decision.
To introduce interpretability, we turn to EBMs.

EBMs are accurate, inherently interpretable predictive models that combine the structure of generalized additive models (GAMs) with the training power of modern boosting. 
GAMs are an extension of linear and generalized linear models\footnote{Generalized linear models are a broad class of models that extend ordinary linear regression to outcomes that are not well-modeled by a Gaussian with constant variance.}, in which linear terms, such as $\beta_i x_i$,  are replaced with learned 1D \textit{shape functions} $f_i(x_i)$, yielding an additive \textit{score} of the form
\begin{align}
    \mathcal{F}_{\text{GAM}}(x) = \beta_0 + \sum_{i=1}^P f_i(x_i),
\label{eq:GAM}
\end{align}
where $P$ is the number of features~\cite{Hastie1986}.
Since each feature's contribution can be visualized directly by plotting $f_i$, this structure preserves interpretability while still allowing nonlinear relationships.

A limitation of purely additive models is that they cannot capture \textit{interactions}, i.e., situations where the effect of one feature depends on the value of another.
A standard extension is to augment the GAM score with a small number of pairwise interaction terms,
\begin{align}
    \mathcal{F}_{\text{GA$^2$M}}(x) = \beta_0 + \sum_{i=1}^P f_i(x_i) + \sum_{(i,j)\in\mathcal{I}}\!f_{ij}(x_i,x_j),
\label{eq:GA2M}
\end{align}
where $\mathcal{I}$ is a small set of interacting feature pairs and each $f_{ij}$ is a learned two-dimensional \textit{interaction function}.
This form, sometimes called a generalized additive model with pairwise (i.e., second-order) interactions (GA$^2$M, where “2” indicates the highest allowed interaction order), retains interpretability since the model can still be inspected through one-dimensional shape plots $f_i$ and a small number of two-dimensional interaction surfaces $f_{ij}$~\cite{Lou2013}.

Training GAM and GA$^2$M models amounts to learning from data functions $\{f_i\}$, and optionally $\{f_{ij}\}$, that minimize an appropriate loss function.
In principle, one could fit these functions using splines or other smooth function classes; however, in practice, the functions must be flexible enough to capture sharp thresholds and non-smooth structure common in experimental data.
Moreover, fitting many functions simultaneously can lead to different terms competing to explain the same variation, reducing stability and complicating interpretation.

%%%%%%%%%%%%%%%%%%%%%%%%%%%%%%%%%%%%%%%%%%%%
%%%%%%%%%%%%%%%%%%%%%%%%%%%%%%%%%%%%%%%%%%%%

EBMs address these issues by learning the GAM/GA$^2$M form using modern boosting, constructing the functions $\{f_i\}$ and (optionally) $\{f_{ij}\}$ through an iterative error-correction process.
Given a dataset of $N$ labeled examples $\mathcal{D}=\{({\bf X}^{(i)}, y^{(i)})\}_{i=1}^N$, where ${\bf X}^{(i)}\in\mathbb{R}^P$ is a $P$-dimensional feature vector (in our case, PIE-derived fit parameters and descriptors) and $y^{(i)}$ is the corresponding class label, we want to learn the score function $\mathcal{F}(x)$ of the GAM/GA$^2$M.
The additive, linear sum of feature contributions is connected with the labels via the so-called \textit{link function} $g(\cdot)$:
\begin{align}
    {\bf y}=g\left(\mathcal{F}({\bf X})\right)
\end{align}
Link functions enable EBM to handle various response types, including classification probabilities, counts, and continuous regressions.
For binary classification, one typically uses the logistic sigmoid,
\begin{align}
    \hat{p}\!\left(y=1\mid {\bf X}\right)=\sigma\!\left(\mathcal{F}({\bf X})\right)
    =\frac{1}{1+e^{-\mathcal{F}({\bf X})}}.
\end{align}
For continuous regression, common choices include the identity link $\mu({\bf X})=\mathcal{F}({\bf X})$ and positive-valued links such as $\mu({\bf X})=\exp(\mathcal{F}({\bf X}))$ (log link) or $\mu({\bf X})=\log\!\left(1+e^{\mathcal{F}({\bf X})}\right)$ (softplus link).
For count data (e.g., Poisson regression), a standard choice is the log link $\mu({\bf X})=\exp(\mathcal{F}({\bf X}))$, which enforces nonnegative predicted mean counts.

The model parameters, i.e., the functions $f_i$ and $f_{ij}$, are fit by minimizing the average loss (also called an \textit{empirical risk}),
\begin{align}
    \mathcal{R}_{\rm emp}(\mathcal{F})=\min_{\mathcal{F}\in\mathcal{G}}\frac{1}{N}\sum_{i=1}^N L\!\left(y^{(i)}, \mathcal{F}({\bf X}^{(i)})\right),
\end{align}
where $\mathcal{G}$ denotes the GAM/GA$^2$M model class and $L$ is a loss function.
For binary classification, a standard choice is the logistic (cross-entropy) loss,
\begin{widetext}
\begin{align}
    L_{\rm log}\!\left(y, \mathcal{F}({\bf X})\right) = -y\log \sigma\!\left(\mathcal{F}({\bf X})\right) -(1-y)\log\!\left(1-\sigma\!\left(\mathcal{F}({\bf X})\right)\right),
\label{eq:logloss}
\end{align}
\end{widetext}
while for regression, a common choice is the squared loss, 
\begin{align}
    L_{\rm sq}(y,\mathcal{F}({\bf X})) = \tfrac12\big(y-\mathcal{F}({\bf X})\big)^2.
\label{eq:sqloss}
\end{align}

Starting from a simple baseline model, the algorithm repeatedly identifies what the current model still gets wrong and adds a small correction that improves the fit.
In practice, residuals are often used in regression problems, and loss gradients are used in classification tasks.
In boosting-based training, the ``error-correction'' step can be made precise: at iteration $t$ one computes pseudo-residuals (negative loss gradients)
\begin{align}
    r_t^{(i)} = -\left.\frac{\partial}{\partial z}\,\ell\!\left(y^{(i)}, z\right)\right|_{z=\mathcal{F}_{t-1}({\bf X}^{(i)})},
\label{eq:pseudores}
\end{align}
and then fits a weak learner to predict $r_t^{(i)}$ from the inputs.
Crucially, each correction is built using a shallow decision tree and the weak learner is constrained to depend only on a single feature (when updating a main effect $f_i$) or a single feature pair (when updating an interaction term $f_{ij}$); the correction is then absorbed into the corresponding $f_i$ or $f_{ij}$.
Thus, each boosting step modifies only one interpretable component of the additive model.
By accumulating many such small, constrained corrections, EBMs can learn highly nonlinear shapes and sharp thresholds while preserving the GAM/GA$^2$M additive structure, ensuring that the final prediction remains fully decomposable into interpretable per-feature and pairwise contributions.
Importantly, EBMs enable not only global explanations in the form of plots of $f_i$ and $f_{ij}$, but also local, per-example breakdowns of contributions without relying on post-hoc interpretability methods.

The same structure that makes EBMs interpretable also limits their utility: EBMs operate on explicit features, so their explanations are only as meaningful as the tabular representation provided to the model.
While many ML methods can learn hierarchical features directly from pixels, EBMs do not ``look'' at raw images in a native way; they instead require semantically meaningful, low-dimensional descriptors whose individual coordinates can be named, plotted, and interpreted.
Fortunately, in the solitonic-excitation setting, such a compact, yet human-intelligible, representation is given by the Mexican hat wavelet fitted to the condensate density profile~\cite{Guo2022}.
This motivates the complementary use of CNNs, which learn feature representations directly from images without requiring a predefined tabular encoding.

%%%%%%%%%%%%%%%%%%%%%%%%%%%%%%%%%%%%%%%%%%%%%%%%%%%%%%%%%%%%%
\subsection{A brief introduction to convolutional neural networks}\label{sec:cnn}
%%%%%%%%%%%%%%%%%%%%%%%%%%%%%%%%%%%%%%%%%%%%%%%%
CNNs are an image-native alternative to EBMs.
Rather than requiring an explicit tabular representation, CNNs learn a hierarchy of spatial features directly.
As an image passes through the layers, a CNN repeatedly transforms it into new arrays called \textit{activations} or \textit{feature maps}.
These feature maps---the model's \textit{internal representation}---are learned intermediate encodings of the input that are useful for the task.
Early layers tend to capture simple local patterns such as edges, blobs, or contrast changes; intermediate layers combine these into higher-level motifs (textures, repeated stripes, localized notches); and deeper layers form task-specific representations that are most informative for separating classes.

This flexibility comes at a price.
In a trained CNN, no single neuron or filter typically corresponds to a single human-meaningful concept; instead, evidence is distributed across many channels and layers, making it difficult to attribute decisions to a small set of interpretable features.

CNNs construct intermediate feature maps by convolving an input image ${\bf X}$ with a bank of learned filters.
Let $\boldsymbol{\mathcal{F}}_0 \equiv {\bf X}$, and denote the output of layer $\ell$ by $\boldsymbol{\mathcal{F}}_{\ell}\in\mathbb{R}^{H_\ell\times W_\ell\times C_\ell}$, where $H_\ell\times W_\ell$ are the spatial dimensions of the feature map and $C_\ell$ is the number of channels.
A typical convolutional block takes the form
\begin{align}
    \boldsymbol{\mathcal{F}}_{\ell+1} = \varphi\!\left({\bf W}_{\ell} * \boldsymbol{\mathcal{F}}_{\ell} + {\bf b}_{\ell}\right),
\label{eq:cnn_layer}
\end{align}
where $*$ denotes convolution, ${\bf W}_{\ell}$ and ${\bf b}_{\ell}$ are learned weights and biases, and, similar to the dense layer in Eq.~\eqref{eq:dense}, $\varphi(\cdot)$ is a pointwise activation function~\cite{Goodfellow2016}.
Common interior activation functions include:

\begin{trailer}{Rectified linear unit (ReLU)}
\vspace{-15pt}
    \begin{align}
        {\rm ReLU}(x)=\max(0,x);
    \end{align}
A common default in CNNs, ReLU is simple, fast, encourages sparse activations, and helps mitigate vanishing gradients compared to saturating nonlinearities.
However, for negative inputs, units sometimes get stuck at 0 (so-called ``dying ReLU''); moreover, because ReLU is unbounded above, large activations can occur.
\end{trailer}

\begin{trailer}{Leaky ReLU}
\vspace{-15pt}
    \begin{align}
    {\rm LReLU}(x)=\max(\alpha x, x),\;\;{\rm where}\;\;0<\alpha\ll 1;
    \end{align}
Leaky ReLU introduces a small negative slope to reduce dying-ReLU behavior while retaining ReLU-like simplicity.
The trade-off is a new hyperparameter (the negative slope); the output remains unbounded above.
\end{trailer}

\begin{trailer}{Gaussian error linear unit (GELU)}
\vspace{-15pt}
    \begin{align}
    {\rm GELU}(x)=x\,\Phi(x),
    \end{align}
where
    \begin{align}
    \Phi(x)=\int_{-\infty}^{x}\tfrac{1}{\sqrt{2\pi}}\exp(-t^2/2)\,dt.
    \end{align}    
GELU is a smooth nonlinearity widely used in Transformer-based models. 
It preserves a nonzero gradient for negative inputs and can improve optimization in some settings, at the cost of slightly higher computational complexity (the exact form involves the Gaussian cumulative distribution function and is often approximated).
\end{trailer}

In addition to convolution, CNN architectures commonly include \textit{pooling} operations that reduce the spatial dimensions of feature maps while retaining the most important information.
Pooling acts independently on each channel and replaces small spatial neighborhoods with a single summary value.
For example, in max pooling with a window size $k\times k$,
\begin{align}
    \mathcal{F}_{\ell+1}(m,n,c) = \max_{\substack{i=1,\dots,k\\ j=1,\dots,k}} \mathcal{F}_{\ell}(km+i, kn+j, c),
\end{align}
where $m$ and $n$ denote the vertical and horizontal spatial indices, respectively, and $c$ indexes the channel, reduces the spatial resolution by a factor of $k$ in each direction.
By progressively decreasing spatial resolution, pooling helps CNNs build representations that are more robust to small translations and reduces the computational cost of deeper layers.
In modern architectures, pooling is sometimes replaced or supplemented by strided convolutions that achieve a similar downsampling effect.

After several convolutional and pooling layers, the feature maps $\boldsymbol{\mathcal{F}}_{\ell}$ encode increasingly abstract, spatially compressed representations of the input.
To convert these representations into a prediction, the network first aggregates the spatial information into a compact feature vector.
This is often done using \textit{global pooling}, which averages or maximizes each channel across its spatial dimensions,
\begin{align}
    {\bf h}({\bf X})_c = \frac{1}{H_\ell W_\ell} \sum_{u=1}^{H_\ell} \sum_{v=1}^{W_\ell} \mathcal{F}_{\ell}(m,n,c),
\end{align}
producing a feature vector ${\bf h}({\bf X})\in\mathbb{R}^{C_\ell}$.
This vector summarizes the presence and strength of the learned features across the entire image.

For classification tasks, a final linear layer maps this feature vector to a set of output scores, or \textit{logits},
\begin{align}
    {\bf z}({\bf X}) = {\bf W}_{\rm out}\,{\bf h}({\bf X}) + {\bf b}_{\rm out}.
\end{align}
These logits are then converted into predictions using an appropriate \textit{output activation function}:

\begin{trailer}{sigmoid}
\vspace{-15pt}
\begin{align}
    Pr(y=1|{\bf X}) = \sigma\!\left(z({\bf X})\right) = \left[1+\exp\left(-z({\bf X})\right)\right]^{-1};
\end{align}
The sigmoid is used for binary classification and maps a scalar logit to a value between $0$ and $1$.
\end{trailer}

\begin{trailer}{softmax}
\begin{align}
    Pr(y=k|{\bf X}) &= {\rm softmax}({\bf z}({\bf X}))_k \notag \\
    & = \frac{\exp\left(z_k({\bf X})\right)}{\sum_{j=1}^{K}\exp\left(z_j({\bf X})\right)};
\end{align}
The softmax is used for multi-class classification; it converts the logit vector into a probability distribution over $K$ mutually exclusive classes.
\end{trailer}

\begin{trailer}{linear (identity)}
\begin{align}
    I({\bf X}) = {\bf X}.
\end{align}
The trivial identity activation function is often used in regression tasks to predict continuous numerical values directly.
\end{trailer}

Thus, CNNs map the input image ${\bf X}$ to a prediction through a sequence of learned feature transformations followed by a task-appropriate output link.

Given a labeled dataset $\mathcal{D}=\{({\bf X}^{(i)}, y^{(i)})\}_{i=1}^N$, the network parameters are learned by minimizing an empirical loss function.
For classification tasks, a common choice is the cross-entropy loss,
\begin{align}
    L_{\rm ce}(\boldsymbol{\theta}) = -\frac{1}{N} \sum_{i=1}^{N} \log Pr_{\boldsymbol{\theta}} \!\left(y^{(i)} \mid {\bf X}^{(i)}\right),
\end{align}
where $\boldsymbol{\theta}$ denotes all learned weights and biases.
Optimization algorithms such as stochastic gradient descent adjust the parameters to minimize this loss, enabling the network to learn feature representations and classification rules directly from the image data.
However, unlike EBMs, where the learned functions correspond directly to named input features, the features learned by CNNs are implicit and must be interpreted indirectly.
Post-hoc tools such as gradient-based saliency maps~\cite{Simonyan2013} or Grad-CAM~\cite{Selvaraju2020} can highlight pixels that influence the output, but these methods provide heuristic visualizations rather than the exact additive decomposition available in EBMs.
We use CNNs as a high-performance baseline for comparing EBM performance and interpretability.

%%%%%%%%%%%%%%%%%%%%%%%%%%%%%%%%%%%%%%%%%%%%%%%%%%%%%%%%%%%%%
\subsection{Results: classification benchmarking}
%%%%%%%%%%%%%%%%%%%%%%%%%%%%%%%%%%%%%%%%%%%%%%%%
Each image in the reduced soliton dataset is represented as a 2D array ${\bf X}{\in}\mathbb{R}^{H\times W}$, with $H{=}164$ and $W{=}132$, and the corresponding Mexican hat fit parameters introduced in Sect.~\ref{sec:soliton_dataset} stored as a 1D array ${\bf Y}{\in}\mathbb{R}^{P}$, with $P{=}14$.
In the context of supervised ML, where each data point is accompanied by a corresponding label, we represent the dataset as $\mathcal{D} = \{({\bf X}^{(i)},{\bf Y}^{(i)}, y^{(i)})\}_{i=1}^N$, with labels $y^{(i)}{\in}\{0,1,2\}$ indicating excitation type (longitudinal soliton, partial, and vortex, respectively).

The workflow for EBM-based classification progresses in two stages.
The first uses the reduced soliton dataset as defined in Sect.~\ref{sec:problem_setup} and serves as a baseline for assessing how well an interpretable, physics-informed model can separate longitudinal solitons, partial excitations, and vortices.
The second stage uses the same reduced dataset and augments the vortex class with physically meaningful, symmetry-preserving transformations to test whether improved representation of the rarest class leads to better classification performance.
These two stages allow us to separate the baseline capability of the EBM from the gains achievable through targeted augmentation of the most underrepresented excitation type.

The EBM performs strongly in both settings, and targeted augmentation leads to a clear improvement in aggregate performance in terms of both accuracy and the F1 score.
Precision measures how often predictions assigned to a given class are correct; recall measures how often true examples of that class are successfully identified; and the F1-score summarizes the balance between precision and recall by taking their harmonic mean, which penalizes large discrepancies between the two; as a result, the score is high only when both precision and recall are simultaneously high.
The balanced accuracy increases from $92.2~\%$ on the baseline reduced dataset to $94.3~\%$ after vortex augmentation, while the average F1-score rises from $94.0~\%$ to $96.2~\%$.
Taken together, these results indicate that the main limitation of the baseline model is not a lack of expressive power, but the scarcity of representative vortex examples in the training data.

%%%%%%%%%%%%%%%%%%%%%%%%%%%%%%%%%%%%%%%%%%%%%%%%
\begin{table}[t]
\caption{Classification report for the reduced soliton dataset, comparing baseline performance of the EBM trained on the reduced dataset (Base.) with performance obtained after applying physically meaningful augmentations to the vortex class (Augm.). 
Metrics are grouped by soliton class.
}
\label{tab:ebm_perf}
\centering
\begin{ruledtabular}
\begin{tabular}{lcccccc}
Metric & \multicolumn{2}{c}{Longitudinal} & \multicolumn{2}{c}{Partial} & \multicolumn{2}{c}{Vortex} \\
& Base. & Augm. & Base. & Augm. & Base. & Augm. \\
\hline
Precision & 0.98 & 0.93 & 0.92 & 0.97 & 0.95 & 0.96 \\
Recall    & 0.97 & 0.95 & 0.85 & 0.98 & 0.92 & 0.93 \\
F1-score  & 0.97 & 0.94 & 0.88 & 0.98 & 0.94 & 0.94 \\
\hline
Support   & 446  & 160  & 13   & 446  & 159  & 27   \\
\hline
ROC-AUC   & 0.997 & 0.986 & 0.936 & 0.997 & 0.981 & 0.976 \\
PR-AUC    & 0.993 & 0.994 & 0.998 & 0.994 & 0.993 & 0.999 \\
\end{tabular}
\end{ruledtabular}
\end{table}
%%%%%%%%%%%%%%%%%%%%%%%%%%%%%%%%%%%%%%%%%%%%%%%%

Detailed quantitative results are summarized in Table~\ref{tab:ebm_perf}.
For each case, we report the standard per-class classification metrics: precision, recall, F1-score, and support (the number of test examples in the class).
We also report one-vs-rest\footnote{In the one-vs-rest setting, each class is treated as the positive class in turn, with all remaining classes grouped as negatives.} ROC-AUC (receiver operating characteristic--area under the curve) and PR-AUC (precision-recall--area under the curve), which quantify how well the model separates the target class from all others across classification thresholds.

On the baseline reduced dataset, the EBM performs strongly on the two majority classes, achieving F1-scores of $0.97$ for longitudinal solitons and $0.94$ for partial excitations.
Performance on the vortex class is also encouraging, but noticeably weaker, with precision $0.92$, recall $0.85$, and F1-score $0.88$.
This behavior is consistent with the dataset's strong class imbalance: while the model learns reliable decision boundaries for the dominant classes, it has fewer examples from which to infer a stable representation of vortices.

Applying physically meaningful augmentation to the vortex class leads to clear improvements across all vortex metrics, with recall increasing from $0.85$ to $0.93$, and the F1-score from $0.88$ to $0.95$.
At the same time, performance on the longitudinal and partial classes remains essentially unchanged, indicating that augmentation enhances minority-class recognition without degrading performance on the majority classes.
These results suggest that the EBM is not fundamentally limited by its expressive power on this task; rather, a significant portion of the remaining error can be attributed to the scarcity of representative vortex examples in the training data.
This overall improvement is driven almost entirely by better identification of the vortex class, rather than by changes in the already strong performance on the two majority classes.

%%%%%%%%%%%%%%%%%%%%%%%%%%%%%%%%%%%%%%%%%%%%
\begin{figure*}[t]
\begin{minipage}[t]{0.60\linewidth}
\vspace{0pt}
    \includegraphics{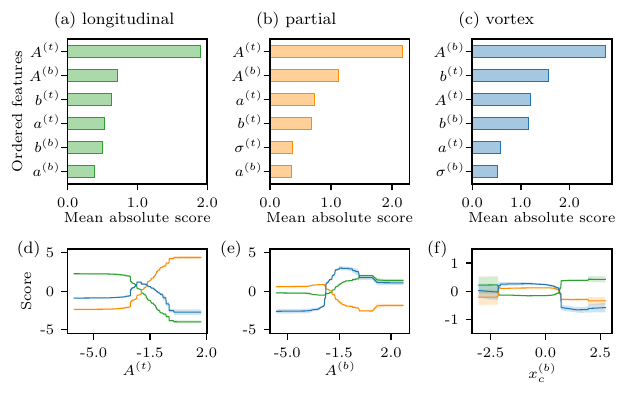}
\end{minipage}
\hspace{0.02\linewidth}
\begin{minipage}[t]{0.33\linewidth}
\vspace{-10pt}
    \caption{
    EBM results.
    Top row: global feature importance, quantified by the mean absolute contribution of each feature to the model score for (a) longitudinal solitons, (b) partial excitations, and (c) vortices.
    The ordering highlights which parameters are most influential in distinguishing each excitation type.
    Bottom row: representative feature curves for (d) $A^{(t)}$, (e) $A^{(b)}$, and (f) $x_c^{(b)}$, illustrating how variations in individual parameters affect the model score.
    The vertical axis (``Score'') represents the contribution of the corresponding feature to the model output: positive values increase the likelihood of the class, while negative values decrease it.
    The three colors in panels (d)--(f) correspond to the longitudinal solitons (green), partial excitations (orange), and vortices (green) labels, respectively.
    }
    \label{fig:ebm_features}
\end{minipage}
\end{figure*}
%%%%%%%%%%%%%%%%%%%%%%%%%%%%%%%%%%%%%%%%%%%%

A key advantage of EBMs is that their decision process can be inspected directly in terms of the features used for training.
Figure~\ref{fig:ebm_features} shows both global and local aspects of this interpretability: the top row ranks features by mean absolute score, indicating which fitted parameters matter most for each class decision, while the bottom row shows representative shape functions that quantify how changes in a given feature shift the model score, where the vertical axis corresponds to the additive contribution of the feature to the class-specific logit (score). 
For this analysis separate Mexican-hat fits were performed to density profiles from the top and bottom portions of the image (extending the single-profile fits in Fig.~\ref{fig:pie_params}); the resulting fit parameters are labeled by $(t)$ and $(b)$.
For all three classes, the most prominent features are the excitation amplitudes $A^{(t/b)}$ together with the shoulder parameters $a^{(t/b)}$ and $b^{(t/b)}$, indicating that the model relies primarily on the depth and asymmetry of the density depletion when distinguishing excitation types.

The first notable difference is in the global feature-importance rankings in Fig.~\ref{fig:ebm_features}a--c.
For the longitudinal and partial classes, the most important parameters are $A^{(t)}$, $a^{(t)}$, and $b^{(t)}$, indicating that classification is driven primarily by the amplitude and detailed shape of the excitation in the top part of the condensate, together with the $A^{(b)}$ parameter capturing the vertical asymmetry in the condensate.
By contrast, the vortex class is dominated by $A^{(b)}$, $b^{(t)}$, $A^{(t)}$, and $b^{(b)}$, showing that vortex identification relies most strongly on the combination of excitation amplitude and asymmetric shoulder structure.
In the language of Fig.~\ref{fig:pie_params}c, this suggests that vortices are distinguished by the interplay between depletion depth and asymmetric distortion, consistent with their more complex density signatures.

The shape-function in Fig.~\ref{fig:ebm_features}d--e provides a particularly clear interpretation of the model's decision logic.
The unequal contributions of $A^{(t)}$ and $A^{(b)}$ reflect the underlying bias in the dataset---with only top partial solitons---capturing the directional asymmetry in the excitation.
Longitudinal solitons exhibit approximately symmetric density depletions, whereas partial excitations and vortices are intrinsically asymmetric.
Consequently, the relative imbalance between $A^{(t)}$ and $A^{(b)}$ provides a key discriminative signal, and the model correctly assigns different contributions to these features to capture this structure.
As shown in Fig.~\ref{fig:ebm_features}d--e, increasing $A^{(t)}$ strongly favors the partial-excitation class while suppressing the longitudinal class, whereas increasing $A^{(b)}$ favors the longitudinal and, to some extent, vortex class, while penalizing partial excitations.
Importantly, the model does not rely on these amplitudes independently, but rather on their combined and nonlinear contributions, allowing it to distinguish symmetric from asymmetric excitation profiles.

In contrast to the amplitude-based features, the positional feature $x_c^{(b)}$ exhibits a nearly flat contribution across the range of interest, indicating that it plays a negligible role in classification.
This is also consistent with the underlying physics: while solitonic excitations can appear at different locations within the condensate, their classification is determined primarily by their shape and symmetry rather than their absolute position.
Thus, the EBM correctly identifies $x_c^{(b)}$ as a non-discriminative feature in this setting, further supporting the alignment between the learned model and physical intuition.

We tested two CNN-based classification architectures. 
The first is a simple CNN with three convolutional layers followed by a fully connected layer established a baseline for image-native learning on the reduced soliton dataset.
The second is a more advanced architecture specifically designed for solitonic-excitation classification, adapted from Ref.~\cite{Guo2021}.
These two configurations allow us to assess how architectural choices impact performance.
Since the models used in this chapter are not hyperparameter-optimized, all experiments are performed with a fixed training set and a held-out test set, without a separate validation set.

%%%%%%%%%%%%%%%%%%%%%%%%%%%%%%%%%%%%%%%%%%%%%%%%
\begin{table}[t]
\caption{Classification report for CNN-based models on the reduced soliton dataset, comparing the mean performance over $10$ runs for a simple CNN architecture (Simp.)  with the best observed performance for the CNN architecture adapted from Ref.~\cite{Guo2021} (Mod.).
Metrics are grouped by soliton class.}
\label{tab:cnn_perf}
\centering

\begin{ruledtabular}
\begin{tabular}{lcccccc}
Metric & \multicolumn{2}{c}{Longitudinal} & \multicolumn{2}{c}{Partial} & \multicolumn{2}{c}{Vortex} \\
& Simp. & Mod. & Simp. & Mod. & Simp. & Mod. \\
\hline
Precision & 0.92(1) & 0.94 & 0.83(4) & 0.86 & 0.90(8) & 0.79 \\
Recall    & 0.96(1) & 0.96 & 0.77(3) & 0.80 & 0.62(4) & 0.85 \\
F1-score  & 0.94(1) & 0.95 & 0.80(2) & 0.83 & 0.73(2) & 0.82 \\
\hline
Support   & 446 & 446 & 159 & 159 & 27 & 27 \\
\hline
ROC-AUC   & 0.955(5) & 0.953 & 0.942(6) & 0.943 & 0.984(5) & 0.983 \\
PR-AUC    & 0.972(2) & 0.969 & 0.888(13) & 0.894 & 0.824(24) & 0.819 \\
\end{tabular}
\end{ruledtabular}
\end{table}
%%%%%%%%%%%%%%%%%%%%%%%%%%%%%%%%%%%%%%%%%%%%%%%%

For the simple CNN, we report average performance across $10$ independent training runs to account for variability from random initialization and stochastic optimization.
The model achieves a mean balanced accuracy of $78.3(1.3)~\%$ and an F1-score of $0.82(1)$.
Unsurprisingly, performance is strongest for the longitudinal class, with recall $0.96(1)$ and F1-score $0.94(1)$, and moderate for partial excitations, with recall $0.77(3)$ and F1-score $0.80$.
The vortex class remains the most challenging, with a recall $0.62(4)$ and an F1-score $0.73(2)$.
This behavior reflects the dataset's class imbalance: the network learns robust representations for the dominant classes but struggles to generalize to the relatively scarce vortex examples.

The architecture adapted from Ref.~\cite{Guo2021} delivers substantially improved performance.
The best observed model achieves a balanced accuracy of $87.0~\%$ and an F1-score of $0.87$.
Both minority classes benefit from the improved architecture, with particularly notable gains for the vortex class, whose recall increases from $0.62(4)$ to $0.85$ and F1-score from $0.73(2)$ to $0.82$.
These results indicate that, when provided with sufficient architectural capacity and appropriate inductive bias, CNNs can learn discriminative representations directly from image data, even for underrepresented classes.

Despite these improvements, the performance of the CNN models is consistently below that of the EBM-based approach for every metric considered, including ROC-AUC and PR-AUC, with the largest advantage in PR-AUC for the minority classes.
The uniformly high values of the ROC-AUC and PR-AUC for all classes, as shown in Table~\ref{tab:cnn_perf}, indicate that the model assigns well-separated scores to each excitation type even before a specific decision threshold is chosen.
The improvement in vortex ROC-AUC from $0.936$ to $0.976$ after augmentation shows that the augmented EBM ranks vortex examples more reliably relative to non-vortex examples, consistent with the observed increase in vortex recall and F1-score.
However, the lower PR-AUC for vortices, about $0.82$ regardless of the CNN architecture, shows that this separability does not translate into equally strong precision--recall behavior for the rare vortex class.

This gap highlights that, for this dataset, the physics-informed feature representation provides a more effective basis for classification than learning directly from pixels alone.
In some sense, identifying the feature representation is part of the overall architecture of an EBM.
While CNN performance would likely improve with substantially larger labeled datasets, achieving such improvements would require targeted data collection and significant additional labeling effort, especially for the rarest excitation types.
Moreover, CNN's internal decision rules are not directly accessible.
Saliency maps~\cite{Simonyan2013} and Grad-CAM~\cite{Selvaraju2020} can highlight pixels influential for a given prediction, but they do not provide an explicit, low-dimensional model ofwhich physical cues drive the decision or how different cues trade off.
The EBM, by contrast, incorporates domain knowledge through its feature representation, reducing the reliance on large labeled datasets and enabling strong performance in a data-constrained setting.

%%%%%%%%%%%%%%%%%%%%%%%%%%%%%%%%%%%%%%%%%%%%%%%%%%%%%%%%%%%%%
\section{Discussion and outlook}
%%%%%%%%%%%%%%%%%%%%%%%%%%%%%%%%%%%%%%%%%%%%%%%%%%%%%%%%%%%%%
A central message of this chapter is that the effectiveness of machine learning methods in scientific settings depends critically on the quality, structure, and size of the underlying dataset.
In both examples considered here, even relatively large datasets benefited from carefully designed augmentation strategies that improved coverage of relevant physical features.
Equally important is the use of validation---and ideally independent test-sets---to ensure that reported performance reflects genuine generalization rather than overfitting.

Our first example of reference reconstruction illustrates that relatively simple linear-algebra-based methods can perform competitively while remaining interpretable.
In these ``models'', the learned components can be directly visualized as images, providing insight into correlated variations in the data.
More complex deep learning approaches yielded only modest performance gains, at the expense of interpretability.
Our second example, soliton classification, provides an even stronger contrast: the physics-informed, interpretable model not only matched but exceeded the performance of the more complex CNN-based approach.

So, ``can machine learning be explainable?'' is a resounding {\it yes}.
Results in this section demonstrate that machine learning models can be both accurate and interpretable, and that the commonly assumed ``tradeoff'' between performance and explainability need not hold in all settings.
In particular, when domain knowledge can be incorporated into the representation---as in the case of physics-informed features---interpretable models may achieve strong performance even in data-limited regimes.

At the same time, it is important to recognize that different modeling approaches scale differently with data.
Deep learning methods are likely to benefit more strongly from large, well-labeled datasets, whereas interpretable, feature-based models can achieve robust performance with fewer labeled examples by leveraging prior knowledge but may require tabular representations.
In practice, the choice of model therefore depends not only on achievable accuracy, but also on the availability of labeled data, the cost of data curation, and the need for interpretability and validation.
In such cases, you, the scientist, must make a judgment call weighing the relative value of raw performance against understanding the machine learning decision process, accepting that, on average, high-performing black-box solutions often lead to unexpected outcomes in seemingly random situations.

Advancing machine learning for scientific applications will require continued attention to dataset design, feature representation, and evaluation protocols (including choosing appropriate loss functions and determining relevant success measures) that emphasize not only performance but also reliability, interpretability, and consistency with underlying physical principles.
Such considerations will be essential for establishing trust in machine learning systems deployed in scientific discovery and measurement-driven domains.
%

%%%%%%%%%%%%%%%%%%%%%%%%%%%%%%%%%%%%%%%%%%%%%%%%%%%%%%%%%%%%%
\begin{acknowledgments}
IBS thanks E.~Gvozdiovas, A.~M.~Pi\~{n}eiro and M.~Zhao for assistance in assembling datasets for the reference recovery section.
This work was partially supported by the National Institute of Standards and Technology.
The views and conclusions contained in this paper are those of the authors and should not be interpreted as representing the official policies, either expressed or implied, of the U.S. Government. 
The U.S. Government is authorized to reproduce and distribute reprints for Government purposes notwithstanding any copyright noted herein. 
Any mention of commercial products is for information only; it does not imply recommendation or endorsement by NIST.
\end{acknowledgments}
%%%%%%%%%%%%%%%%%%%%%%%%%%%%%%%%%%%%%%%%%%%%%%%%%%%%%%%%%%%%%

%%%%%%%%%%%%%%%%%%%%%%%%%%%%%%%%%%%%%%%%%%%%%%%%%%%%%%%%%%%%%

%%%%%%%%%%%%%%%%%%%%%%%%%%%%%%%%%%%

%%%%%%%%%%%%%%%%%%%%%%%%%%%%%%%%%%%%%%%%%%%%%%%%%%%%%%%%%%%%%
\end{document}